\documentclass[a4paper,11pt]{article}
\usepackage{jinstpub} % for details on the use of the package, please see the JINST-author-manual
\usepackage{graphicx}% Include figure files
\usepackage{dcolumn}% Align table columns on decimal point
\usepackage{bm}% bold math
\usepackage{subfigure,dcolumn}
\usepackage{hyperref}% add hypertext capabilities
\usepackage{pgfplots}
\usepackage{booktabs}
\usepackage{multirow}
\title{High-Accuracy Schottky Diagnostics for Low-SNR Betatron Tune Measurement in Ramping Synchrotrons}

\author[a,b]{P. Sun}
\author[a,c,1]{M. Zhang\note{Corresponding author.}}
\author[a,c]{R. Yuan}
\author[a]{D. Li}
\author[c]{J. Dong}
\author[a,b]{and Y. Shi}

\affiliation[a]{Shanghai Institute of Applied Physics, Chinese Academy of Sciences,\\
Shanghai 201800, China}
\affiliation[b]{University of Chinese Academy of Sciences,\\
China}
\affiliation[c]{Shanghai Advanced Research Institute, Chinese Academy of Sciences,\\
Shanghai 201204, China}

\emailAdd{zhangmanzhou@sinap.ac.cn}

\abstract{This study introduces a novel real-time betatron tune measurement algorithm, utilizing Schottky signals and an FPGA-based backend architecture, specifically designed for rapidly ramping synchrotrons, with particular application to the Shanghai Advanced Proton Therapy (SAPT) facility. The developed algorithm demonstrates improved measurement accuracy under challenging operational conditions, especially in scenarios with limited sampling time and signal-to-noise ratios (SNR) as low as \(-20\) dB. By applying Short-Time Fourier Transform (STFT) analysis, the algorithm effectively accommodates the rapid increase in revolution frequency from 4 MHz to 7.5 MHz over 0.35 seconds, along with tune shifts. A macro-particle simulation methodology is employed to generate Schottky signals, which are then combined with real noise collected from SAPT without beam to simulate practical conditions. The proposed betatron tune measurement algorithm integrates advanced spectral processing techniques and an enhanced peak detection algorithm specifically tailored for low SNR conditions. Simulation validation confirms the superior performance of the proposed algorithm over conventional approaches in terms of measurement accuracy, stability, and system robustness, while meeting the stringent operational requirements of proton therapy applications. This innovative approach effectively addresses critical limitations associated with Schottky diagnostics for betatron tune measurement in rapidly ramping synchrotrons operating under low SNR conditions, laying a robust foundation and providing a viable solution for advanced applications in proton therapy and related accelerator physics fields.
}

\keywords{Instrumentation for particle-beam therapy; Instrumentation for particle accelerators and storage rings - high energy (linear accelerators, synchrotrons); Digital signal processing (DSP); Data processing methods}

\arxivnumber{2412.19171} % Only if you have one

\begin{document}
\maketitle
\flushbottom

\section{\label{sec:intro}Introduction}

SAPT is a synchrotron-based proton therapy facility designed, constructed, and reliably operated in Shanghai during its clinical trial. To optimize the third-order resonance slow extraction of the SAPT facility~\cite{zhang2023sapt}, the main ring must be capable of performing betatron tune measurements under different energy and bunching or drifting beam. However, due to the absence of an integrated design for the betatron tune measurement system, the current setup can only measure the tune while bunching. Based on the practical experience of the SAPT facility, the residual oscillations after injection do not seem to be effectively sustained, making it difficult to obtain a coherent tune signal from the beam position monitor (BPM) data. This process involves using a slow extraction kicker for excitation, followed by applying a Fast Fourier Transform (FFT) to signals from the BPM. Additionally, excitation evidently interferes with the extraction process. This limitation hinders the system from meeting the demands of further optimization. Therefore, a feasible option at present is to perform tune measurements by obtaining the incoherent transverse oscillation signal, specifically using the Schottky signal measurements method.

Since the suggestion of the stochastic cooling concept by Simon van der Meer in 1969~\cite{vanderMeer:312939}, Schottky signal measurements have become a widely used non-invasive tool for determining beam properties, including momentum spread, betatron tune, synchrotron frequency, and chromaticity. In practice, achieving a high SNR has consistently proven to be challenging. The SNR of the measured signal is influenced by factors such as the longitudinal length of the pick-up, the sensitivity of the BPM, and the operating frequency of the system. As a result, the design and manufacturing of the pick-up may be constrained by various factors, complicating efforts to ensure a high SNR. Given these limitations, an alternative approach is necessary for measuring the betatron tune at the SAPT facility.

This paper outlines the method and procedure for measuring the betatron tune under conditions of low SNR and fluctuating betatron tune. We incorporate simulated signals—both with and without coherent components—at varying revolution frequencies and betatron tunes, combined with real-world noise collected from an ADC, to evaluate the reliability and general applicability of this method. Depending on signal quality and specific analytical requirements, various spectral smoothing algorithms can be implemented to optimize the balance between computational efficiency, measurement precision, and noise suppression capabilities.

The paper is organized as follows: 
Section~\ref{sec:simulation} presents the theoretical foundations of betatron tune measurement using Schottky diagnostics, alongside methodologies for time-domain beam dynamics-based Schottky signal simulation.
Section~\ref{sec:algorithm} introduces a key innovation of this paper: an enhanced peak-detection algorithm specifically designed for synchrotrons operating under low SNR conditions during bunching or drifting. The method begins with a detailed data preprocessing procedure, including data acquisition, STFT window length selection, and coherent signal exclusion. This is followed by spectral processing, which involves signal smoothing to reduce noise interference, followed by spectral mapping to ensure that components from different harmonic regions are accurately aligned and summed. Advanced signal processing techniques are then integrated to enhance robustness, including Exponential Moving Average (EMA) for reference tune estimation based on historical power spectral densities (PSDs), an online median filter for shot noise reduction, Kalman filtering for multi-sensor fusion, and Weighted Linear Combination (WLC) for more accurate and reliable tune measurement. These steps collectively enable effective identification of the betatron tune under challenging conditions.
In Section~\ref{sec:experiments}, we validate the general applicability of the proposed method under various operational scenarios. The performance of the proposed algorithm is evaluated using multiple metrics and compared with that of the conventional peak-detection algorithm. The results demonstrate that the proposed algorithm achieves superior performance in terms of accuracy, stability, and robustness under challenging conditions.
Section~\ref{sec:limitations} discusses the limitations and future work related to the validation of the proposed algorithm. 
Finally, concluding remarks are provided in Section~\ref{sec:conclusion}.

\section{\label{sec:simulation}Macro-Particle Simulation}
\subsection{\label{subsec:theorerical_background}Theoretical Background}
\subsubsection{\label{subsubsec:longitudinal_signal}Longitudinal Signal}

The time-domain current signal of a single particle \( i \) circulating in the synchrotron within a coasting beam, as detected by a pickup electrode can be expressed as~\cite{lannoy2024impact, nolden2001instrumentation}:
\begin{align}
I^i(t) &= e \sum_{k=-\infty}^{\infty} \delta(t-\Delta t_i-kT_0)
\end{align}
The longitudinal signal at the \(n\)-th harmonic in a coasting beam can be expressed as
\begin{align}
I^i(t) &=\frac{\omega_0e}{2\pi}e^{jn\omega_0(t-\Delta{t_i})}
\label{eqn:lon_drift}
\end{align}
In a bunched beam configuration, particles execute synchrotron oscillations. The signal at the \(n\)-th harmonic can be mathematically expressed as~\cite{betz2017bunched, VanderMeerDiagnostics, lasocha2020lon, lannoy2022lhc, Lasocha:2900915}:

\begin{align}
I^i(t) &=\frac{\omega_0 e }{2\pi} \mathfrak{R}\big(\sum_{p=-\infty}^{\infty} J_p\left( n \omega_0 \hat{\tau}_i \right) e^{j \left( n \omega_0 t + p \Omega_{s_i} t + p \psi_i \right)}\big) \label{eqn:lon_bessel}
\end{align}
where:
\begin{itemize}
    \item \( e \) is the elementary charge,
    \item \( \delta \) is the Dirac delta function,
    \item \( \Delta t_i \) is the arrival time difference between particle \( i \) and the synchronous particle,
    \item \( T_0 = \frac{1}{f_0} \) is the nominal revolution period,
    \item \( \omega_0 = 2\pi f_0 \) is the angular nominal revolution frequency in radians,
    \item \( \hat{\tau}_i \) is the synchrotron oscillation amplitude of particle \( i \),
    \item \( \Omega_{s_i} = 2\pi f_{s_i} \) is the angular synchrotron frequency of particle \( i \),
    \item \( \psi_i \) is the initial synchrotron phase, and
    \item \( J_p \) is the Bessel function of the first kind of order \( p \).
\end{itemize}

The nominal bunch length of SAPT is about one-quarter of the circumference of the ring according to the RF voltage. The synchrotron frequency of particle $i$ depends solely on the synchrotron oscillation amplitude and follows from the solution of the pendulum equation~\cite{lasocha2020lon}, yielding:

\begin{align}
    \Omega_{s_i} = \frac{\pi}{2K\left(\sin\left(\frac{\omega_\text{RF} \hat{\tau}_i}{2}\right)\right)} \Omega_{s_0} \label{eqn:synchrotron_frequency}
\end{align}
where $K([0, 1]) \to \left[\frac{\pi}{2}, \infty\right)$ is the complete elliptic integral of the first kind, $\omega_\text{RF} = 1 \cdot \omega_0$ is the RF frequency, and $\Omega_{s_0} = q_s \cdot \omega_0$ is the zero-amplitude synchrotron frequency, where $q_s$ represents the synchrotron tune. The initial synchrotron phase of particle $i$, $\psi_i$, is drawn from a uniform distribution over the range $(-\pi, \pi)$. 
% The longitudinal shape of ${10^11}$ protons power spectral density (PSD) under SAPT conditions is shown in Fig.~\ref{fig:mc_lon}.

% \begin{figure}[h]
%     \includegraphics[width=0.7\hsize]{MacroParticlePSDLongitudinal.png}
%     \caption{Simulated longitudinal Schottky spectrum of $10^{11}$ protons at the 5th harmonic of the revolution frequency at $7.5 \, \text{MHz}$.}
%     \label{fig:mc_lon}
% \end{figure}

\subsubsection{\label{subsubsec:transverse_signal}Transverse Signal}

The transverse Schottky signal spectra are derived from the dipole moment of the beam. For a single particle in a coasting beam within a proton synchrotron, the transverse dipole Schottky time-domain signal at the \(n\)-th harmonic can be mathematically expressed as~\cite{betz2017bunched, VanderMeerDiagnostics, lasocha2022tran, Lasocha:2900915, lannoy2024impact}:

\begin{align}
    D_{\pm q}^i(t) &\propto \frac{\hat{x}_i}{2} e^{j\left(\left(n \pm q \right)\omega_0\left(t-\Delta t_i\right) + \phi_i \right)}
    \label{eqn::tran_drift}
\end{align}
For a single particle in a bunched beam within a proton synchrotron, the transverse dipole Schottky time-domain signal at the \(n\)-th harmonic can be mathematically expressed as:
\begin{align}
D_{\pm q}^i(t) &\propto \frac{\hat{x}_i}{2} \mathfrak{R} \big( \sum_{p=-\infty}^\infty J_p\big((n \hat{\tau}_i \pm \frac{\hat{Q}_i}{\Omega_{s_i}}) \omega_0\big) \notag \\
&\quad e^{j\big((n\pm Q)\omega_0 t + p \Omega_{s_i} t + p \psi_i + \phi_i\big)}\big), \label{eqn:tran_bessel}
\end{align}
where:
\begin{itemize}
    \item $\hat{x}_i$ is the betatron oscillation amplitude, 
    \item $q$ is the fractional part of the nominal tune $Q$, 
    \item $\hat{Q}_i = Q \xi \frac{\hat{p}_i}{p_0}$ is the amplitude of the tune oscillations~\cite{lasocha2022tran}, which may have any sign,
    \item $\xi$ is the chromaticity, 
    \item $\hat{p}_i$ is the amplitude of momentum oscillation,
    \item $\phi_i$ is the initial betatron phase of particle $i$, which is also drawn from a uniform distribution over the range $(-\pi, \pi)$, similar to $\psi_i$. 
\end{itemize}

\subsection{Beam Dynamics Simulation}

Two primary methodologies exist for performing macro-particle simulations: the Monte Carlo approach and the beam dynamics-based approach.

The Monte Carlo simulation methodology is grounded in theoretical formulations for both longitudinal (Equations~\ref{eqn:lon_drift} and~\ref{eqn:lon_bessel}) and transverse Schottky signals (Equations~\ref{eqn::tran_drift} and~\ref{eqn:tran_bessel}). These approaches depend on analytical expressions, which inherently neglect statistical fluctuations due to finite particle numbers and machine imperfections, including magnetic field errors, RF jitter, and lattice nonlinearities. To overcome these limitations and produce more realistic Schottky signals that better reflect actual beam conditions, we employ beam dynamics simulations using the Xsuite framework. 

\begin{figure*}
    \includegraphics[width=\hsize]{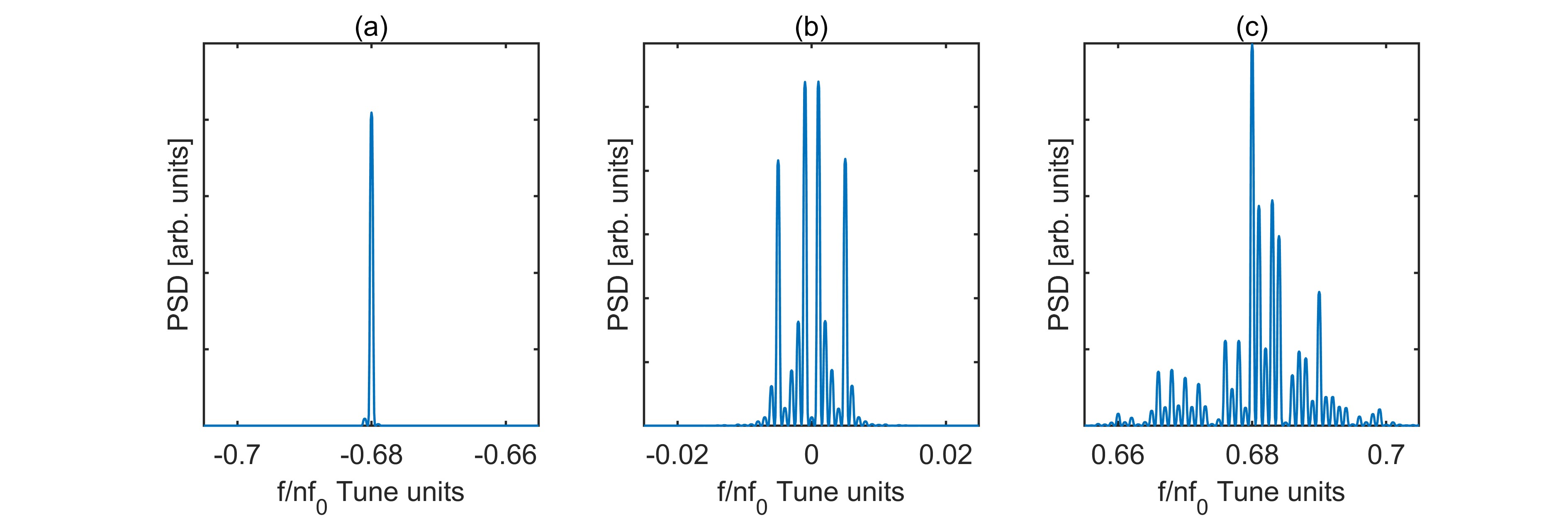}
    \caption{PSDs of the longitudinal (Fig.~\ref{fig:beam_dynamics}(b)) and transverse horizontal spectra (Fig.~\ref{fig:beam_dynamics}(a), Fig.~\ref{fig:beam_dynamics}(c)) are shown for a simulation involving \(10^5\) macro-particles representing \(10^{11}\) protons at the fifth harmonic of the revolution frequency (\(7.5\,\text{MHz}\)). Typical SAPT values, as listed in Table~\ref{tab:synch_params}, are assumed.}
    \label{fig:beam_dynamics}
\end{figure*}

A methodology for synthesizing Schottky spectra from macro-particle simulations utilizing the Xsuite code~\cite{lannoy2022lhc, lannoy2024impact} is implemented. Xsuite comprises a suite of Python packages specifically designed for high-fidelity simulation of beam dynamics in particle accelerators. The framework incorporates modules for generating and manipulating particle ensembles, along with capabilities for precise single-particle tracking. Through application of characteristic SAPT parameters, Schottky spectra at specific energies and harmonics are computed, as demonstrated in Fig.~\ref{fig:beam_dynamics}.

The spectra generated using Xsuite exhibit random fluctuations in the Schottky signal. Furthermore, the configurable sampling frequencies in the \texttt{xtrack.BeamPositionMonitor} class allow for continuous acquisition of transverse beam centroid positions (x and y), effectively producing waveforms sampled at the designated frequency. This method more closely reflects the manner in which BPM data is typically acquired in operational accelerators. Higher sampling frequencies enable observation of transverse beam oscillations at higher harmonics, which is crucial for transverse Schottky diagnostics. The PSDs of the lower and upper sidebands are denoted by \(P_{-T}\) and \(P_{+T}\), respectively.

\begin{table}
\centering
\caption{\label{tab:synch_params} Key parameters of the SAPT facility.}
\begin{tabular}{ll}
\toprule
\textbf{Parameter} & \textbf{Value} \\
\midrule
Circumference & 24.6\,m \\
Intensity & \( 1 \times 10^{11} \) protons per bunch \\
Injection Energy & 7\,MeV \\ 
Extraction Energy & 70--235\,MeV \\ 
Tune (Injection) & \( Q_x = 1.7, \quad Q_y = 1.45 \) \\
Tune (Extraction) & \( Q_x = 1.68, \quad Q_y = 1.40 \) \\
Chromaticities & \( \xi_x = -1.46, \quad \xi_y = -1.34 \) \\
\( \epsilon_x, \epsilon_y \) & \( 2\pi \) mm$\cdot$mrad \\ 
\( q_s \) & 0.001 \\
\( \alpha \) & 0.3175 \\
Transition Gamma & \(\gamma_t = 1.576\) \\
RF Voltage & 1500\,V \\
\( h_{rf} \) & 1 \\
\bottomrule
\end{tabular}
\end{table}

\section{\label{sec:algorithm}Tune Measurement}

This section presents the comprehensive methodology for accurate betatron tune measurement. Initially, data acquired from the BPM undergoes preprocessing operations, as detailed in Section~\ref{subsec:preprocessing}. Subsequently, the preprocessed data undergoes spectral processing, as described in Section~\ref{subsec:spectral_processing}. Following spectral processing, an enhanced peak-detection algorithm is implemented to achieve accurate and stable measurement results, as elaborated in Section~\ref{subsec:epda}. The measurement results subsequently undergo post-processing procedures, as presented in Section~\ref{subsec:postprocessing}. Finally, the detailed implementation architecture for the tune measurement system is presented in Section~\ref{subsec:workflow}.

\subsection{Data Preprocessing\label{subsec:preprocessing}}

\subsubsection{Data Acquisition}

A resonant stripline BPM~\cite{keil2010commissioning, petersen2017fermilab, citterio2009design, dehler2005resonant, kesselman2001resonant} is currently under development for the detection of Schottky signals at the SAPT facility. The device features a bandwidth of 3 MHz, centered at approximately 36 MHz, to optimize spectral coverage. The detected signal undergoes sequential processing: it first passes through a low-pass filter to remove frequency components above the upper bound of the BPM, then through a hybrid to suppress the common-mode longitudinal component, followed by a band-pass filter to reject out-of-band frequencies, and is ultimately digitized by an ADC.

Data acquisition can be implemented using either a variable or fixed sampling rate, with minimal differences in preprocessing procedures. Two primary methodologies are considered:

\textbf{Method 1:} ADC triggering via a harmonic of the revolution frequency, obtained through phase-locking an external generator to the radio frequency (RF) signal. This approach mitigates the time dependence of the betatron frequency, ensuring the transverse Schottky sideband appears in a consistent spectral region, thereby simplifying spectral processing and tune measurement.

\textbf{Method 2:} Implementation of oversampling and application of the bandpass sampling theorem. In an ideal ADC, quantization noise is modeled as white noise uniformly distributed across the frequency range from DC to half the sampling rate. The total quantization noise power, denoted as \(\sigma_q^2\), is given by:
\begin{align}
    \sigma_q^2 = \frac{\Delta^2}{12},
\end{align}
where \(\Delta\) denotes the quantization step size. For an N-bit quantizer with full-scale range \(A\), the step size is given by \(\Delta = \frac{A}{2^N}\). The corresponding PSD is:
\begin{align}
    \text{PSD}_{q} = \frac{\Delta^2}{12f_\text{sampling}},
\end{align}
where \(f_\text{sampling}\) denotes the ADC sampling rate. This expression demonstrates that an increase in \(f_\text{sampling}\) results in a proportional decrease in the quantization noise PSD, \(\text{PSD}_q\).

For a bandpass signal, employing a sampling frequency significantly higher than twice the upper bound of the signal spectrum, followed by lowpass or bandpass filtering and subsequent decimation, can enhance both the SNR and the effective number of bits (ENOB) of the ADC, thereby compensating for its limited physical resolution. The oversampling ratio is defined as \(\text{OSR} = f_\text{sampling}/(2BW_\text{BPM})\), where \(BW_\text{BPM}\) denotes the bandwidth of the BPM. The corresponding improvement in SNR due to oversampling and filtering can be expressed as:
\begin{align}
    \Delta \text{SNR}_{\text{OS}} = 10 \log_{10}(\text{OSR}) \, \text{dB}
\end{align}
The corresponding improvement in ENOB is given by:
\begin{align}
    \Delta_\text{ENOB} = \frac{10\log_{10} \text{OSR}}{6.02}.
\end{align}

The decimation ratio is determined based on the constraints of bandpass sampling, which allow for sampling at rates significantly below the conventional Nyquist limit~\cite{lyons1997understanding}. The allowable range of the bandpass sampling frequency is mathematically defined as:
\begin{align}
    \frac{2f_H}{m+1} &\leq f_\text{BP} \leq \frac{2f_L}{m},
\end{align}
where \(f_L\) and \(f_H\) represent the lower and upper bounds of the BPM bandwidth, respectively, \(f_\text{BP}\) denotes the bandpass sampling rate, and \( m = \left\lfloor \frac{f_L}{f_H - f_L} \right\rfloor \).

Under these conditions, the signal spectrum will lie within the frequency interval \((f_1, f_2)\), where \(f_2 \leq f_\text{BP}/2\). To eliminate the time dependence of the betatron frequency, the frequency axis will be divided by $f_0$, normalized to tune units, and processed as described in Section~\ref{subsec:spectral_processing}. This method achieves a similar effect to the first approach but offers greater implementation flexibility.

Practical constraints arise from the synchrotron's ownership by Ruijin Hospital and its continuous clinical operation. Modifications to the synchrotron structure are limited during BPM development. Notably, the second method provides a plug-and-play solution that can be integrated into the existing synchrotron without interfering with the RF cavity or other critical electronics. Consequently, this paper adopts the second methodology.

\subsubsection{STFT Window Length Selection}

\begin{figure}[h]
    \centering
    \includegraphics[width=0.5\hsize]{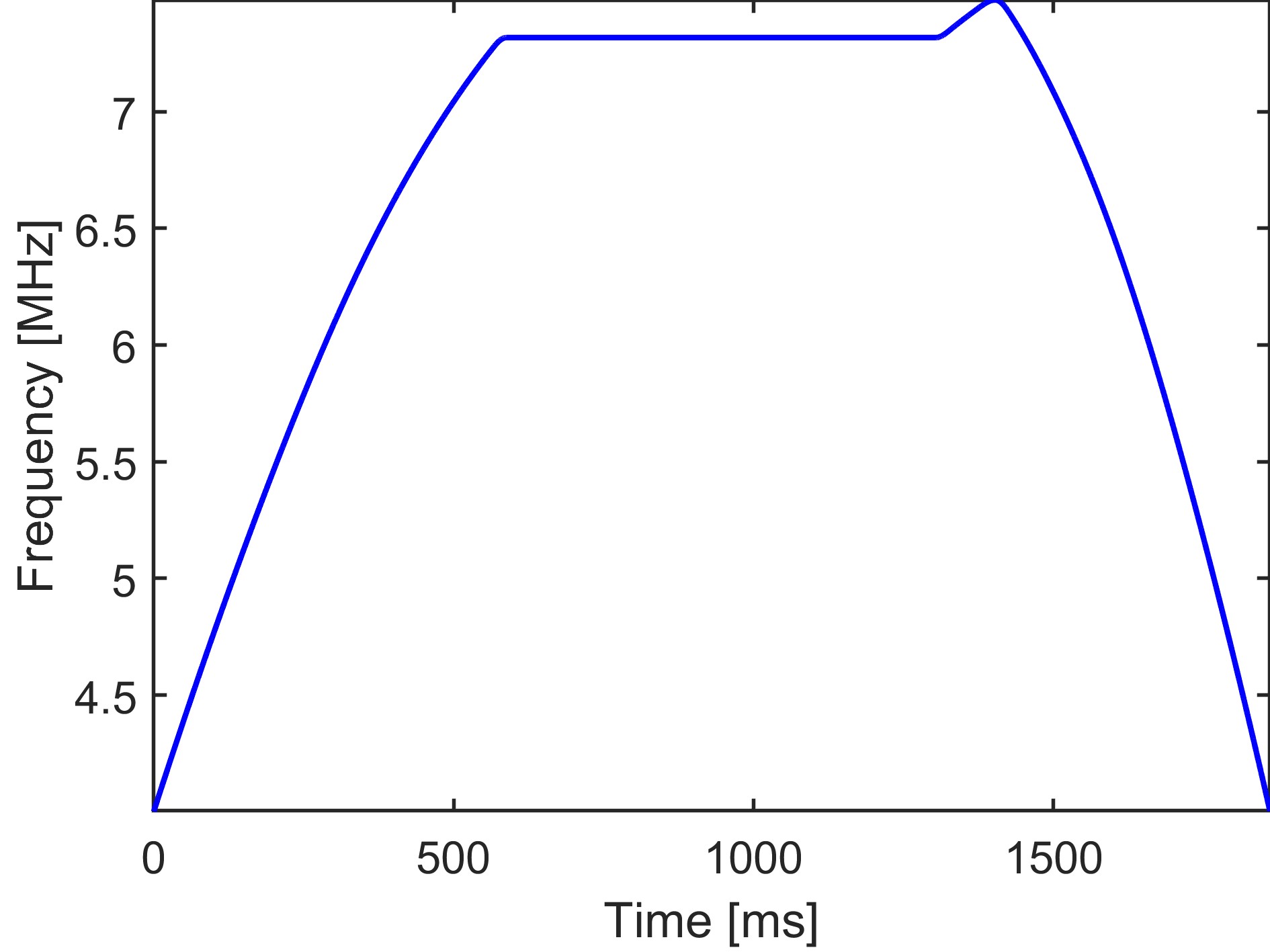}
    \caption{Revolution frequency change during ramping and extraction process.}
    \label{fig:ramping}
\end{figure}

During operation, the proton kinetic energy ranges from 70 MeV to 235 MeV, corresponding to a revolution frequency range of 4 MHz to 7.5 MHz, depending on the tumor depth. The variation in revolution frequency during ramping to maximum energy is shown in Fig.~\ref{fig:ramping}. The system is designed to measure the betatron tune across different energy levels and under both bunched and drifting beam conditions. Additionally, it must be capable of performing tune measurements during both the ramping and extraction procedure, which requires adaptation to revolution frequency variations at a maximum rate of 10 MHz/s. This constraint directly impacts the sampling time; when sampling duration becomes excessively prolonged, sideband locations undergo significant shifts, ultimately compromising both the precision and accuracy of the measured tune values. In the context of SAPT, a 10 kHz variation in revolution frequency during the tune measurement procedure is considered tolerable. Consequently, the STFT window length should not exceed 1 millisecond. Moreover, during tune measurements, the revolution frequency is acquired in real time from other BPMs, thereby further reducing the errors and inaccuracies caused by its variation.

\subsubsection{Coherent Signal Exclusion\label{subsubsec:coherent_exclusion}}

In practical measurements, interference from coherent signals may be encountered, which can distort the desired signal. To mitigate these disturbances, the Root Mean Square (RMS) fit method is commonly employed. This approach aims to minimize the impact of noise by fitting a model to the data that best represents the underlying signal, while reducing the effect of random fluctuations.

The RMS fit procedure involves the following steps:
\begin{enumerate}
    \item \textbf{Data Preparation}: Prepare the data by ensuring it has the desired length and format for processing.
    \item \textbf{Model Selection}: Choose an initial model that approximates the expected signal. This could be a sine wave, a Gaussian function, or another appropriate mathematical representation. In the context of this paper, a sine wave of the form $A\sin(2\pi f x + \phi) $ is selected.
    \item \textbf{RMS Calculation}: The RMS value is computed as the square root of the mean of the squared differences between the model and the measured data. This step quantifies the goodness of the fit between the model and the data.
    \item \textbf{Error Minimization}: The fitting process aims to minimize the RMS error by adjusting the parameters of the model (e.g., amplitude, phase, frequency) until the difference between the model and the measured data is minimized.
    \item \textbf{Residual Evaluation}: After the fitting procedure, the difference between the fitted model and the measured data is evaluated to ensure that the noise component has been adequately removed and that the desired signal is accurately represented.
\end{enumerate}

\begin{figure}[h]
    \centering
    \includegraphics[width=0.5\hsize]{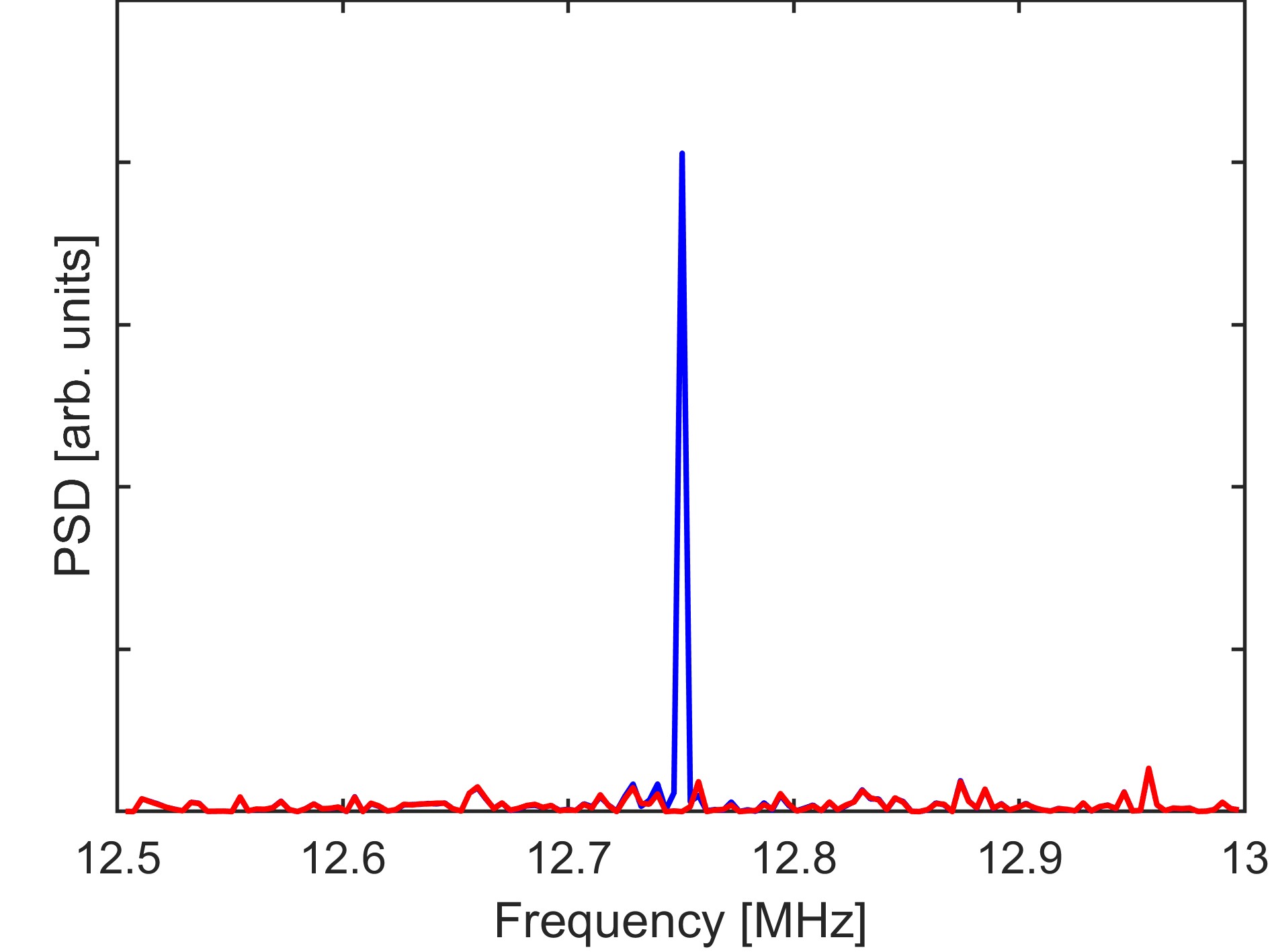}
    \caption{Frequency-domain representation of the signal before and after the exclusion of coherent signal using the RMS fit method. The blue curve represents the original signal spectrum, while the red curve shows the spectrum with coherent signal removed.}
    \label{fig:coherent_exclusion}
\end{figure}

By minimizing the RMS error, this method effectively mitigates the impact of noise, resulting in a more accurate representation of the signal, which is particularly beneficial in the analysis of Schottky diagnostics and signal preprocessing. An example of the exclusion of coherent component from a signal using an RMS fit is presented in Fig.~\ref{fig:coherent_exclusion}. 
% In practice, the coherent component in the Schottky signal can also be subtracted using this method.

\subsection{\label{subsec:spectral_processing}Spectral Processing}

Following the data preprocessing outlined in Section~\ref{subsec:preprocessing}, spectral processing techniques are employed to reduce the impact of noise power. This step is crucial to prevent spectrum degradation, facilitate accurate identification of the transverse Schottky signal's position, and enhance the SNR. Two key components of the spectral processing procedure, namely smoothing and mapping, are discussed in detail in the following subsections.

\subsubsection{\label{subsubsec:smoothing}Smoothing}

The time window for the STFT, comprising $N_t$ data points, is partitioned into $\lceil N_t/N_b \rceil$ batches, where $N_b$ denotes the number of frequency bins per batch. Subsequently, the PSD of each batch, denoted by $P_{b_i}$, is computed. The PSD of each batch undergoes interpolation when higher precision is required, and summation, followed by averaging to yield $\overline{P}$.

To extract the transverse signal spectrum obscured by noise, Gaussian filtering is applied to \( \overline{P} \) using a window of appropriate size.

When selecting the window size for filtering, manual selection reduces the method’s generalizability and level of automation. The implementation of a fixed-length window produces diverse effects contingent upon the frequency resolution employed. When operating at low resolution or with extended window lengths, this approach can result in excessive smoothing of the complete transverse signal profile. Therefore, the window size is determined based on the number of points in the spectrum that encompass the transverse signal spectrum.

The process begins with calculating the width of the transverse signal spectrum. For an unbunched beam, the transverse sideband width is given by  
\begin{align}
\Delta f_{\pm T} &= f_0 \frac{\Delta p}{p} \big|(n \pm q)\eta \pm Q\xi\big|,
\end{align}  
where $\eta$ is the slip factor, and $q$ is the fractional part of the betatron tune $Q$.  

In the case of a bunched beam, the spectrum of a single particle splits into an infinite series of synchrotron satellites spaced by the synchrotron frequency $f_s$. From Equations~\ref{eqn:lon_bessel} and~\ref{eqn:tran_bessel}, we obtain the terms $J_p\left( n \omega_0 \hat{\tau}_i \right)$ and $J_p\big((n \hat{\tau}_i \pm \frac{\hat{Q}_i}{\Omega_{s_i}}) \omega_0\big)$. Since $J_p(x) \approx 0$ for $p > x$, the maximum bandwidth of the transverse spectrum of a single particle is given by~\cite{boussard1986schottky}:  
\begin{align}
\text{BW}_{\pm T} &= 2\omega_0 \big|n\hat{\tau} \pm \frac{\hat{Q}}{\Omega_s}\big| \Omega_s.
\end{align}  
Therefore, the approximate width of the transverse signal spectrum at the $n$-th harmonic is  
\begin{align}
\text{BW} &= \text{BW}_{\pm T} + \Delta f_{\pm T} \nonumber \\
&= 2\omega_0 \big|n\hat{\tau} \pm \frac{\hat{Q}}{\Omega_s}\big| \Omega_s + f_0 \frac{\Delta p}{p} \big|(n \pm q)\eta \pm Q\xi\big|.
\end{align}

The maximum bandwidth of the transverse signal spectrum, denoted as \(\text{BW}_T = \max(\text{BW})\), is computed and selected as the global width to minimize redundant calculations. Given the frequency resolution \(\Delta f\), the number of spectral points encompassing the transverse Schottky signal is determined as:
\begin{align}
    N_T = \left\lceil \frac{\text{BW}_T}{\Delta f} \right\rceil.
\end{align}

Experimental results under varying sampling rates and revolution frequencies indicate that optimal smoothing is achieved when the window size $N_f$ is set to:

\begin{align}
    N_f = \max\big(3, 2\left\lfloor \frac{N_T}{2} \right\rfloor + 1\big).
\end{align}

\begin{figure}
    \centering
    \includegraphics[width=0.5\hsize]{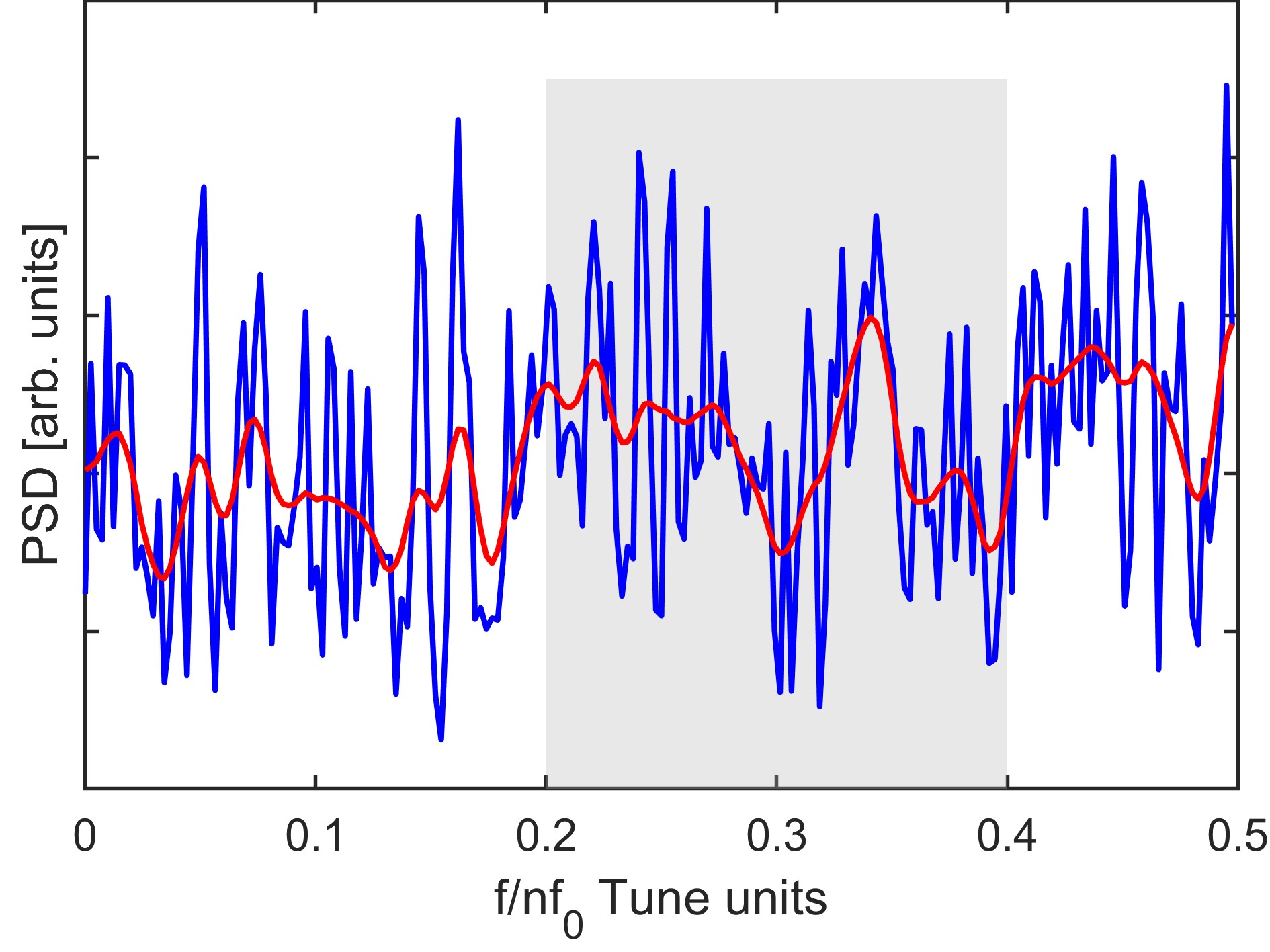}
    \caption{A Gaussian-weighted average with \( \sigma = \frac{N_f - 1}{3} \), encompassing 99.7\% of the energy within \( 3\sigma \), is applied to \( \overline{P} \). The gray-shaded region highlights the frequency range where the transverse sideband is located. The blue line corresponds to the original spectrum, while the red line represents the Gaussian-filtered spectrum.}
    \label{fig:smoothing}
\end{figure}

The determined window size is then applied for smoothing, yielding the red line in Fig.~\ref{fig:smoothing}, denoted as \(P_t\), which represents the PSD at time \(t\).

\subsubsection{\label{subsubsec:folding}Mapping}

\begin{figure}
    \centering
    \includegraphics[width=0.5\hsize]{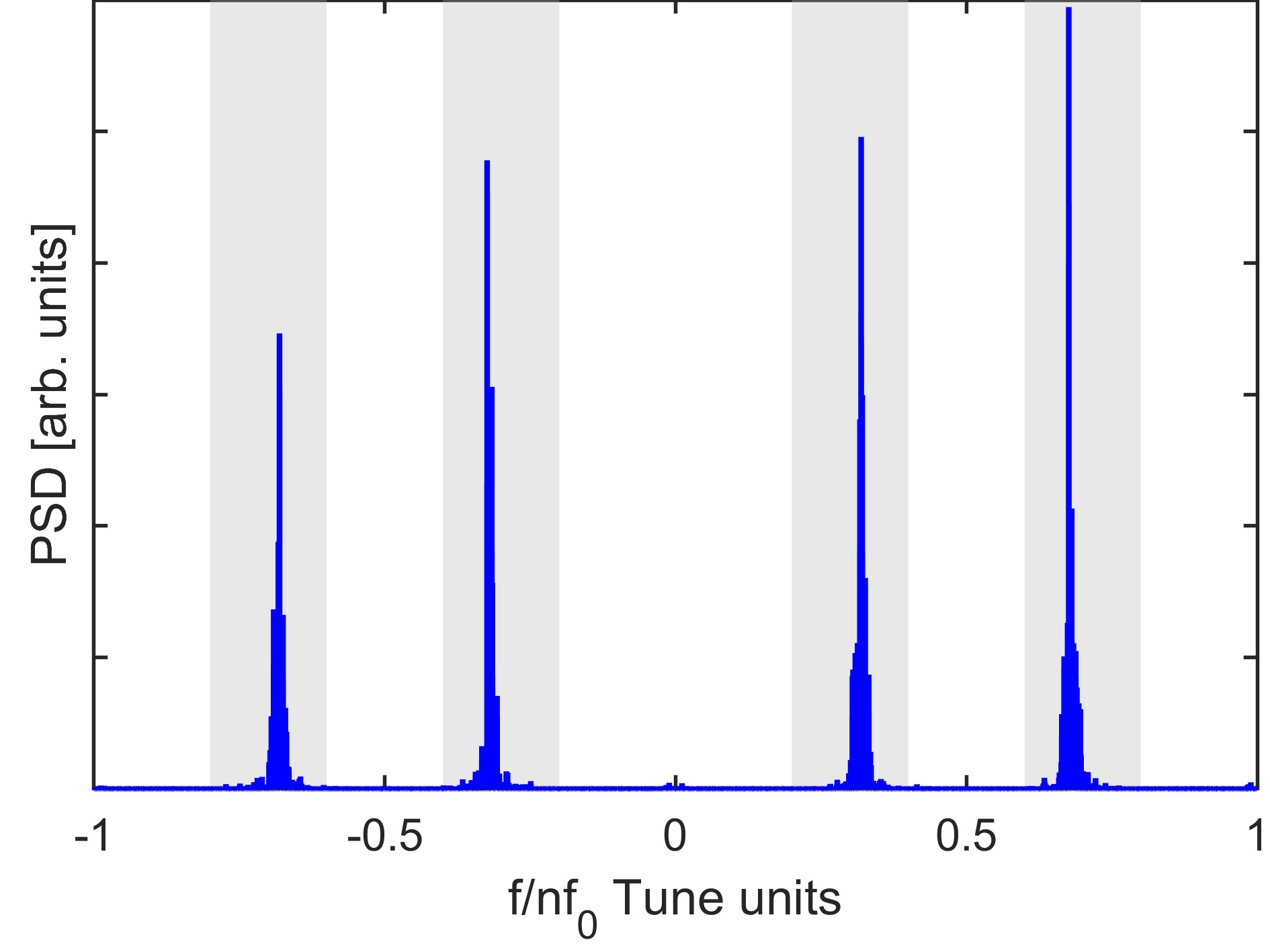}
    \caption{The spectrum contains three revolution harmonics located at -1, 0, and 1, along with four transverse sidebands. The sidebands within the gray-shaded regions are interpolated, inverted if necessary, mapped into the range \( (0, 0.5) \) to improve the SNR.}
    \label{fig:folding}
\end{figure}

In practice, due to the finite bandwidth of the BPM and the choice of sampling rates, multiple sidebands may appear in the spectrum, as shown in Fig.~\ref{fig:folding}. These sidebands can be effectively utilized by mapping them into the range $(0, 0.5)$ in tune units, thereby enhancing the SNR for tune measurements, as demonstrated in~\cite{betz2017bunched}. It is important to note that this technique is specific to tune measurement and does not confer similar advantages for other beam parameter measurements, such as chromaticity. Furthermore, regardless of tune shifts, the betatron tune remains confined to a fixed region in the mapped spectrum, provided that the sampling rate is selected as either an integer harmonic of the revolution frequency or a constant value. 

Since the bandwidth of the BPM under development is 3 MHz and the revolution frequency to be applied ranges from 4 to 7.5 MHz, only one sideband can be detected at most frequencies. Consequently, this procedure serves primarily as a method to eliminate the time dependence of betatron frequency. Nevertheless, when implemented with a wideband detector, this approach offers a viable technique for enhancing the SNR.

\subsection{\label{subsec:epda}Enhanced Peak-Detection Algorithm}

Previous studies have established three primary methodologies for betatron tune identification~\cite{betz2017bunched, lasocha2022tran}. The first methodology, peak detection, identifies the coherent tune by locating the point with the maximum amplitude, which forms the basis of the proposed algorithm.
The second methodology employs spectrum curve fitting, wherein the coherent portion of the sideband is excluded, and an appropriate fitting function is applied to extract the incoherent tune. However, within the SAPT context, data collected from the BPM cannot be guaranteed to possess high SNR and frequency resolution, thereby limiting this methodology's applicability to SAPT. 
The third approach, the Mirrored Difference method, utilizes the symmetry between positive and negative $p$-satellites to determine the precise center of the sidebands. This method, however, proves unsuitable for SAPT applications due to the continuously varying revolution frequency and tune shifting, which precludes the achievement of sufficient frequency resolution required to resolve the internal structure of Bessel satellites.

Experiments in Section~\ref{sec:experiments} demonstrate that the original peak-detection algorithm fails to provide satisfactory generalizability and stability under SAPT conditions, particularly in scenarios with low SNR. To address these limitations, we propose an enhanced peak-detection algorithm designed to improve accuracy and stability in challenging low-SNR environments. The algorithm comprises multiple interconnected components, each playing a distinct role in the tune measurement process. The following subsections present these components in detail. 
% It should be noted that, for the sake of clarity in presentation, all evaluations in this section are based on the assumption that the detector has sufficient bandwidth to detect at least one transverse sideband across all energy regions. Cases where certain energy regions cannot be detected will be discussed in Section~\ref{sec:experiments} to further demonstrate the robustness of the algorithm.

\subsubsection{Exponential Moving Average\label{subsubsec:ema}}

The Exponential Moving Average (EMA) represents a sophisticated statistical technique for smoothing time-series data through an exponentially weighted aggregation of historical observations. In contrast to the Simple Moving Average (SMA), which implements uniform weighting across a data window, the EMA demonstrates enhanced sensitivity by prioritizing more recent measurements. This approach enables the EMA to accumulate PSD contributions from historical data while simultaneously emphasizing contemporary observations.

At a given time \( t \), the current PSD value \( P_t \) is processed, and the updated EMA is formulated as:

\begin{align}
    \text{EMA}_t = \alpha \cdot P_t + (1 - \alpha) \cdot \text{EMA}_{t-1},
\end{align}
where the parameters represent:
\begin{itemize}
    \item \( P_t \): Current power spectral density
    \item \( \text{EMA}_{t-1} \): Preceding EMA value
    \item \( \alpha \): Smoothing factor \( (0 < \alpha \leq 1) \)
\end{itemize}

The smoothing factor \( \alpha \) critically modulates the relative influence of current observations. A diminished \( \alpha \) yields a more attenuated output, whereas an elevated \( \alpha \) accentuates recent dynamical variations. Consequently, the EMA's intrinsic adaptability and computational efficiency render it a prevalent methodology in signal processing, financial analytics, and real-time data filtration.

The position of the global maximum within the exponential moving average \(\text{EMA}_t\) provides the EMA tune measurement. This EMA tune, complemented by the tune acquired using the Weighted Linar Combination (WLC) method (discussed in Section~\ref{subsubsec:wlc}), will undergo online median filtering (detailed in Section~\ref{subsubsec:mf}) to mitigate shot noise interference. These two values represent independent data points collected from discrete sensors, facilitating a comprehensive multi-sensor fusion process.

\begin{figure*}
    \centering
    \includegraphics[width=0.8\hsize]{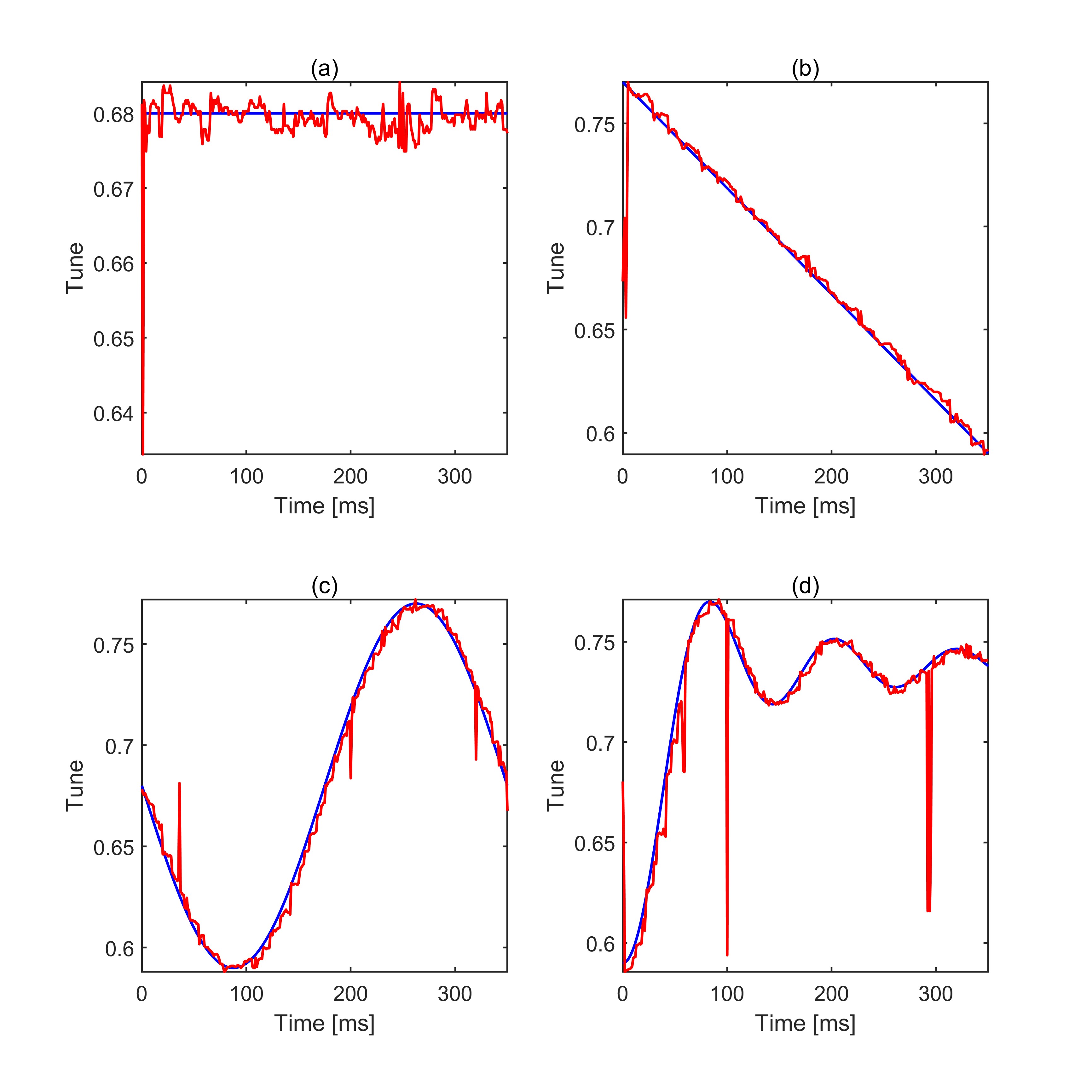}
    \caption{Comparison between the EMA tune (red line), obtained by identifying the global maxima of \(\text{EMA}_t\), and the actual tune (blue line) under different tune shift scenarios. In most cases, the EMA tune remains stable and accurate, albeit with slight latency. However, due to strong background noise, occasional fluctuations in the obtained EMA tunes are observed, which can be attributed to shot noise. The mitigation of these disturbances will be discussed in Section~\ref{subsubsec:mf}. It is important to note that the tune variations depicted in the figures are solely intended to evaluate the accuracy of the acquired reference tune and do not represent the actual tune variations of SAPT during operation.}
    \label{fig:ref_comparison}
\end{figure*}

Given the characteristic stability of betatron tunes in synchrotrons, wherein significant perturbations occur infrequently, the Exponential Moving Average (EMA) method emerges as a robust technique for predicting tune values based on previously acquired PSDs. A comprehensive comparison between the derived EMA tune and the actual tune, based on BPM data obtained from Xsuite simulations, is presented in Fig.~\ref{fig:ref_comparison}.

\subsubsection{Weighted Linear Combination\label{subsubsec:wlc}}

\begin{figure}
    \centering
    \includegraphics[width=0.5\hsize]{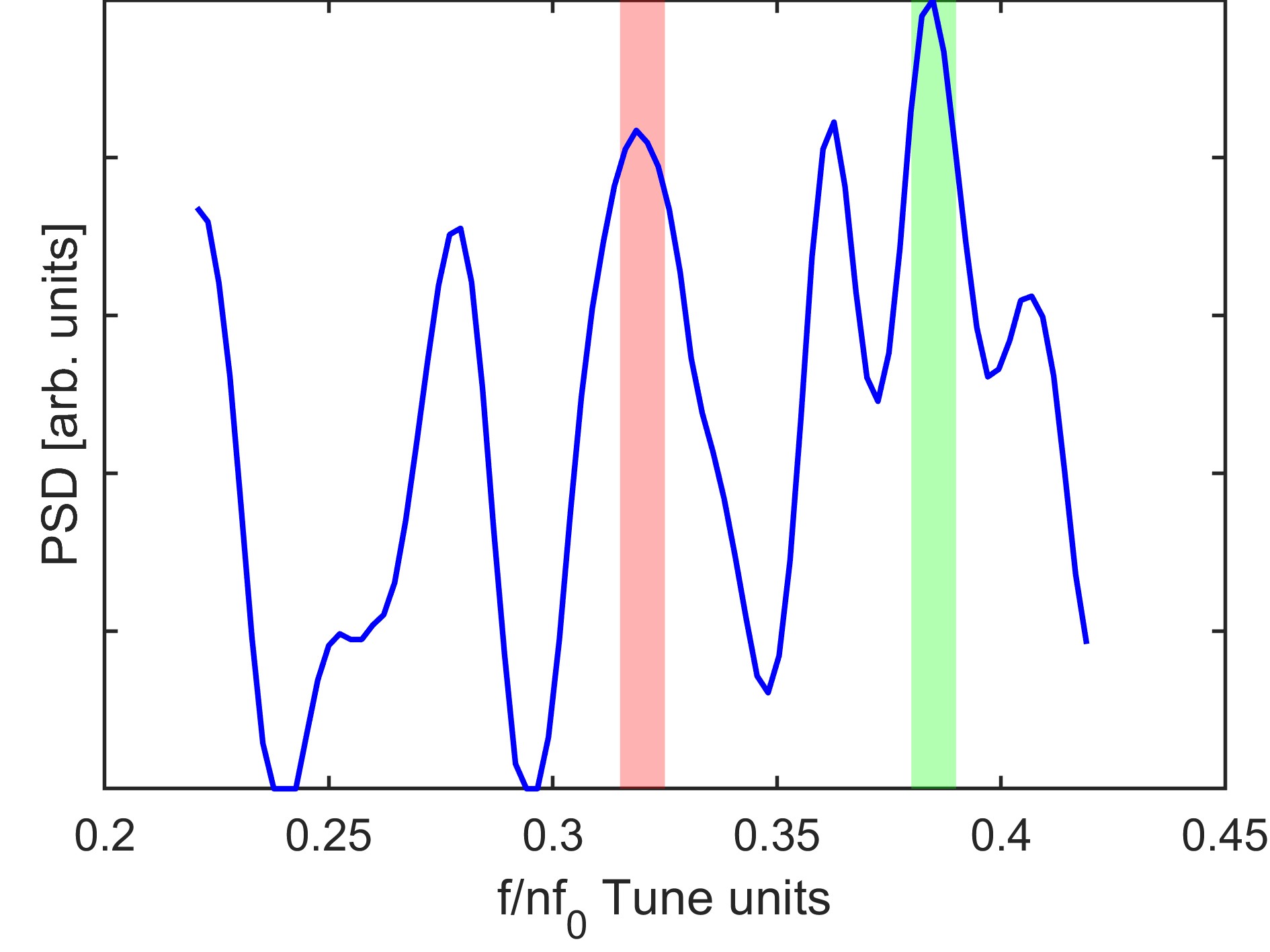}
    \caption{Folded and smoothed spectrum under an SNR of -20 dB with \(q=0.68\). The peak within the red-shaded region represents the location of true betatron tune; however, it does not exhibit the highest amplitude. The original peak-detection algorithm erroneously identifies the peak within the green-shaded region as the measured tune, as it corresponds to the global maximum in the presented spectrum.}
    \label{fig:hard2distinguish}
\end{figure}

The original peak-detection algorithm operates on the premise that the SNR is sufficiently high to render the transverse sideband prominent and readily identifiable, with multiple sidebands expected to fall within the detector bandwidth. Under this assumption, after smoothing, the highest peak within the designated region is selected as the betatron tune. However, this assumption breaks down when the background noise power significantly exceeds the signal power. In the frequency spectrum, this manifests as multiple peaks of similar amplitude, making it difficult to distinguish the true transverse sideband peak, as illustrated in Fig.~\ref{fig:hard2distinguish}.

To leverage the temporal continuity of the betatron tune, Weighted Linear Combination (WLC) is introduced to address the limitations of the conventional peak-detection algorithm. WLC is a mathematical operation that combines multiple variables or signals, each multiplied by a corresponding weight. It is widely used in signal processing, optimization, and data analysis to emphasize or de-emphasize the contribution of specific components. The general form of a weighted linear combination is given by:

\begin{align}
    y = \sum_{i=1}^{n} k_i x_i
\end{align}

where:
\begin{itemize}
    \item \( y \) is the resulting combined value,
    \item \( k_i \) represents the weight assigned to the \( i \)-th component,
    \item \( x_i \) is the \( i \)-th input variable or signal,
    \item \( n \) is the total number of components.
\end{itemize}

The weights $k_i$ are conventionally normalized to satisfy $\sum_{i=1}^{n} k_i = 1$, ensuring that the weighted combination represents a balanced contribution from all constituent inputs. As established in Sections~\ref{subsubsec:ema} and~\ref{subsubsec:mf}, the EMA tune is derived. Given that the betatron tune generally exhibits gradual rather than abrupt variations, we can exploit this characteristic to systematically integrate information from both the previous time step tune and the current time step tune. Consequently, we formulate:
\begin{align}
    q_\text{ref}=w\cdot q_{\text{EMA}_t} + (1-w)\cdot q_{\text{WLC}_{t-1}},\label{eqn:q_ref}
\end{align}
where $q_\text{ref}$ denotes the reference tune for the subsequent WLC procedure, $q_{\text{EMA}_t}$ represents the EMA tune at the current time step, and $q_{\text{WLC}_{t-1}}$ indicates the WLC tune from the previous time step. The implementation of a linear combination for reference tune determination is justified by the observation that when the EMA tune is perturbed by shot noise, it produces a significantly biased reference for the WLC procedure, consequently generating inaccurate WLC tune values. To avoid exclusive dependence on the EMA tune for reference determination, we integrate both the current EMA tune and the previous WLC tune, thereby obtaining a more robust reference with reduced bias.

Before applying WLC, the locations and amplitudes of all local maxima are identified. Subsequently, two factors are considered. The first factor is the distance between each local maximum and filtered reference tune, \( q_\text{ref}\). The distances between all local maxima and \( q_\text{ref}\) are calculated individually, normalized to the range \( [0,1] \), and the corresponding weight is given by  
\begin{align}  
    P_\text{distance} = 1 - \text{distance}.  
\end{align}  
The second factor is the amplitude. The amplitudes of all local maxima are normalized to the range \( [0,1] \), yielding the weight \( P_\text{amplitude} \).  

Next, we introduce a parameter \( k \) to represent the weight of \( P_\text{distance} \), while the weight of \( P_\text{amplitude} \) is given by \( 1-k \). The overall confidence of a local maximum, indicating its likelihood of being the actual tune, is then computed as  
\begin{align}  
    \text{conf} = k \cdot P_\text{distance} + (1-k) \cdot P_\text{amplitude}.  
\end{align}  
The local maximum exhibiting the highest confidence value is designated as the measured tune. The measured tune subsequently undergoes online median filtering before being processed through a Kalman filter to yield more stable and consistent results.

\begin{figure}
    \centering
    \includegraphics[width=0.5\hsize]{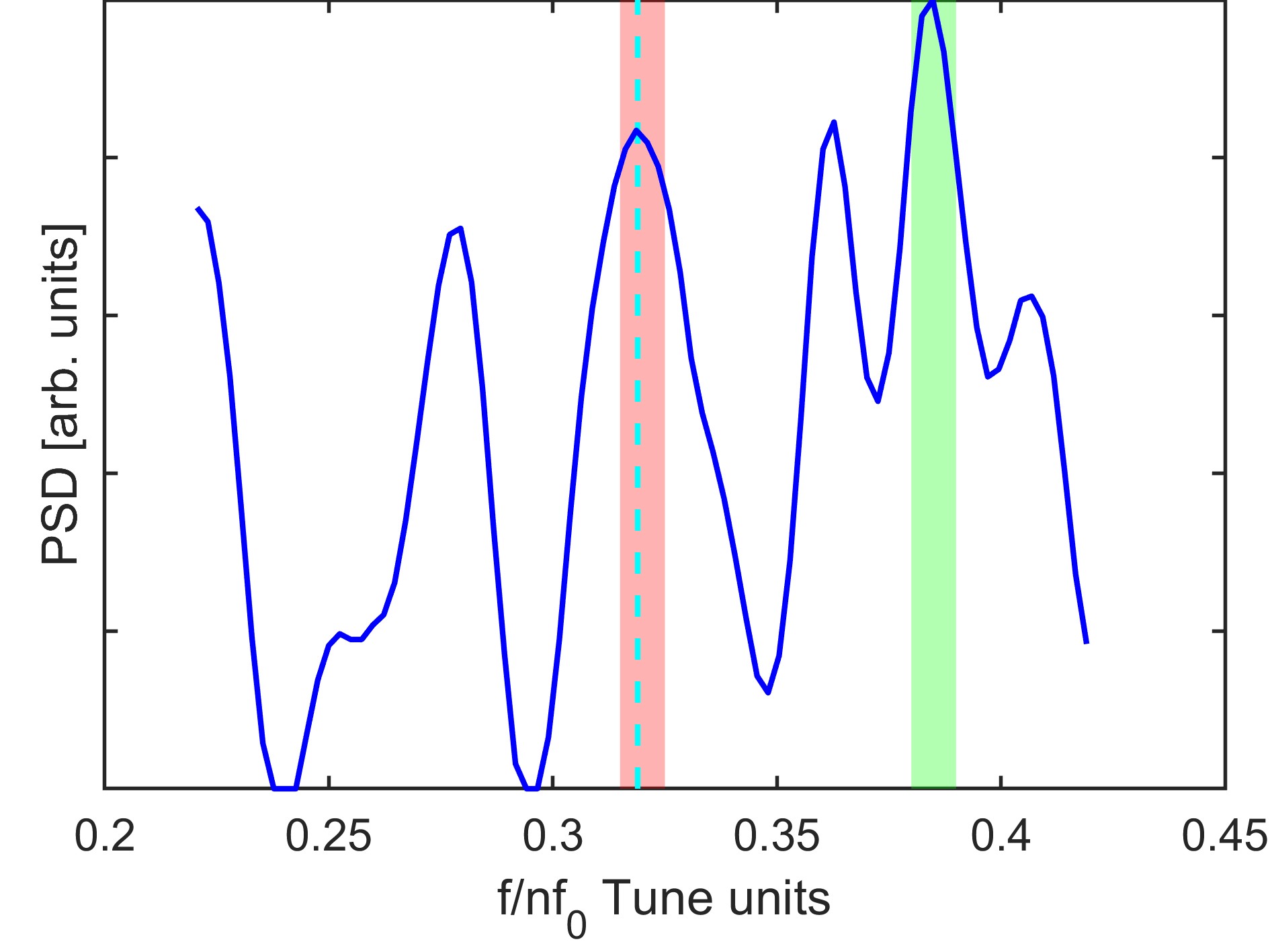}
    \caption{The peak within the red-shaded region represents the true betatron tune. Relying solely on amplitude would erroneously identify the peak within the green-shaded region as the measured tune, as it corresponds to the global maximum in the presented spectrum. The cyan dashed line represents \( q_\text{ref} \). Even though the peak in the red-shaded region is not the global maximum, it still receives significant weight because it is the closest local maximum to the reference tune of the previous time step. After applying WLC, the peak in the red-shaded region has the highest confidence and is thus identified as the WLC tune.}
    \label{fig:wlc}
\end{figure}

WLC leverages the temporal continuity of the tune, assigning greater weight to local maxima that are closer to the tune of the previous time step, rather than relying solely on amplitude. This approach enhances robustness, as shown in Fig.~\ref{fig:wlc}.

\subsubsection{Online Median Filter\label{subsubsec:mf}}

\begin{figure*}
    \centering
    \includegraphics[width=0.8\hsize]{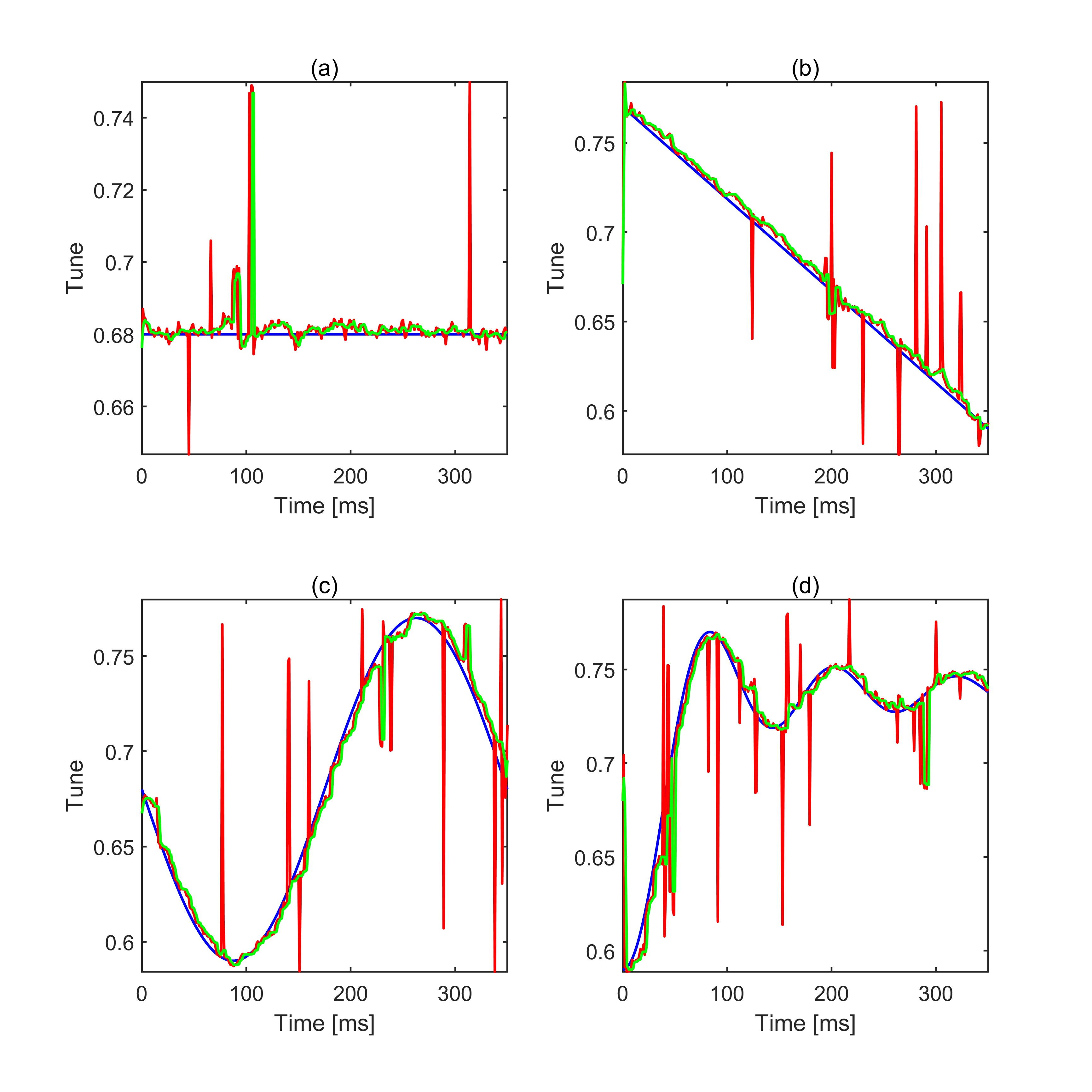}
    \caption{Comparison of the acquired EMA tunes with (green line) and without (red line) online median filtering under different tune shift scenarios. The blue line represents the actual tune. The online median filtering process effectively mitigates shot noise in the results; however, it introduces latency, which depends on the filter's window size.}
    \label{fig:median_comparison}
\end{figure*}

In the proposed enhanced peak-detection algorithm, the EMA tune derived from $\text{EMA}_t$ is processed as sensor data. Typically, the observed values can be characterized as sensor data contaminated by Gaussian white noise. However, occasional observations exhibiting significant deviations from the actual tune can be attributed to shot noise phenomena. To mitigate the effects of sporadic shot noise, an online median filter is implemented. This filter employs a sliding window of optimal size to execute real-time filtering on both the EMA tune and the WLC tune, the latter being processed using the WLC method as detailed in Section~\ref{subsubsec:wlc}. The comparative results before and after application of the online median filter are illustrated in Fig.~\ref{fig:median_comparison}.

\subsubsection{Adaptive Multi-sensor Fusion\label{subsubsec:fusion}}

The adaptive multi-sensor fusion framework employs a Kalman filter to integrate the EMA tune from Section~\ref{subsubsec:ema} and the WLC tune from Section~\ref{subsubsec:wlc}, treating them as inputs from two sensors. The filter dynamically adjusts their contributions based on real-time noise estimation, thereby enhancing the reliability of the state estimate by mitigating the effects of varying measurement noise and occasional shot noise disturbances. The workflow and underlying principles of the adaptive multi-sensor fusion process can be summarized as follows:

\begin{itemize}
    \item \textbf{Initialization:} \\
    The filter initializes with an initial state estimate \(x_0\) and an associated error covariance \(P_0\). The process noise covariance, \(Q\), accounts for system uncertainties, while the measurement noise covariances, \(R_1\) and \(R_2\), are initially assigned equal fixed values. Additionally, a residual history window is maintained for each detector to enable dynamic noise estimation.

    \item \textbf{Prediction Step:} \\
    The prediction step propagates the previous state estimate forward in time under the assumption of an identity state transition model:
    \begin{align}
        x_{\text{pred}} &= x, \\
        P_{\text{pred}} &= P + Q,
    \end{align}
    where \(x_{\text{pred}}\) is the predicted state estimate, and \(P_{\text{pred}}\) is the predicted error covariance incorporating process noise \(Q\).

    \item \textbf{Adaptive Noise Estimation:} \\
    To dynamically adjust for measurement noise, updated noise estimates for each detector are calculated utilizing an exponential smoothing algorithm:
    \begin{align}
        R_i &= \alpha \cdot (z_i - x_{\text{pred}})^2 + (1 - \alpha) R_i, \quad i = 1,2,
    \end{align}
    where $\alpha$ represents the smoothing factor governing the adaptation rate.
    
    \item \textbf{Measurement Fusion:} \\
    Given two independent measurements, \(z_1\) and \(z_2\), obtained from separate detectors with noise variances \(R_1\) and \(R_2\), the normalized weights for each detector are computed as:
    \begin{align}
        w_1 &= \frac{\frac{1}{R_1}}{\frac{1}{R_1} + \frac{1}{R_2}}, \quad
        w_2 = \frac{\frac{1}{R_2}}{\frac{1}{R_1} + \frac{1}{R_2}}.
    \end{align}
    The fused measurement is then obtained as:
    \begin{align}
        z_{\text{fused}} = w_1 z_1 + w_2 z_2,
    \end{align}
    with the corresponding equivalent measurement noise:
    \begin{align}
        R_{\text{fused}} = \frac{1}{\frac{1}{R_1} + \frac{1}{R_2}}.
    \end{align}

    \item \textbf{Update Step:} \\
    The Kalman gain is computed as:
    \begin{align}
        K = \frac{P_{\text{pred}}}{P_{\text{pred}} + R_{\text{fused}}}.
    \end{align}
    The updated state estimate is then given by:
    \begin{align}
        x = x_{\text{pred}} + K \cdot (z_{\text{fused}} - x_{\text{pred}}),
    \end{align}
    and the error covariance is updated as:
    \begin{align}
        P = (1 - K) P_{\text{pred}}.
    \end{align}

    \item \textbf{Process Noise Update:} \\
    The process noise covariance is updated based on the squared magnitude of the fused residual:
    \begin{align}
        Q = \alpha \cdot (z_\text{fused} - x_\text{pred})^2 + (1 - \alpha) Q.
    \end{align}
    This adaptive approach ensures that the filter dynamically responds to variations in both measurement noise and system uncertainties.
\end{itemize}

\begin{figure}
    \centering
    \includegraphics[width=\hsize]{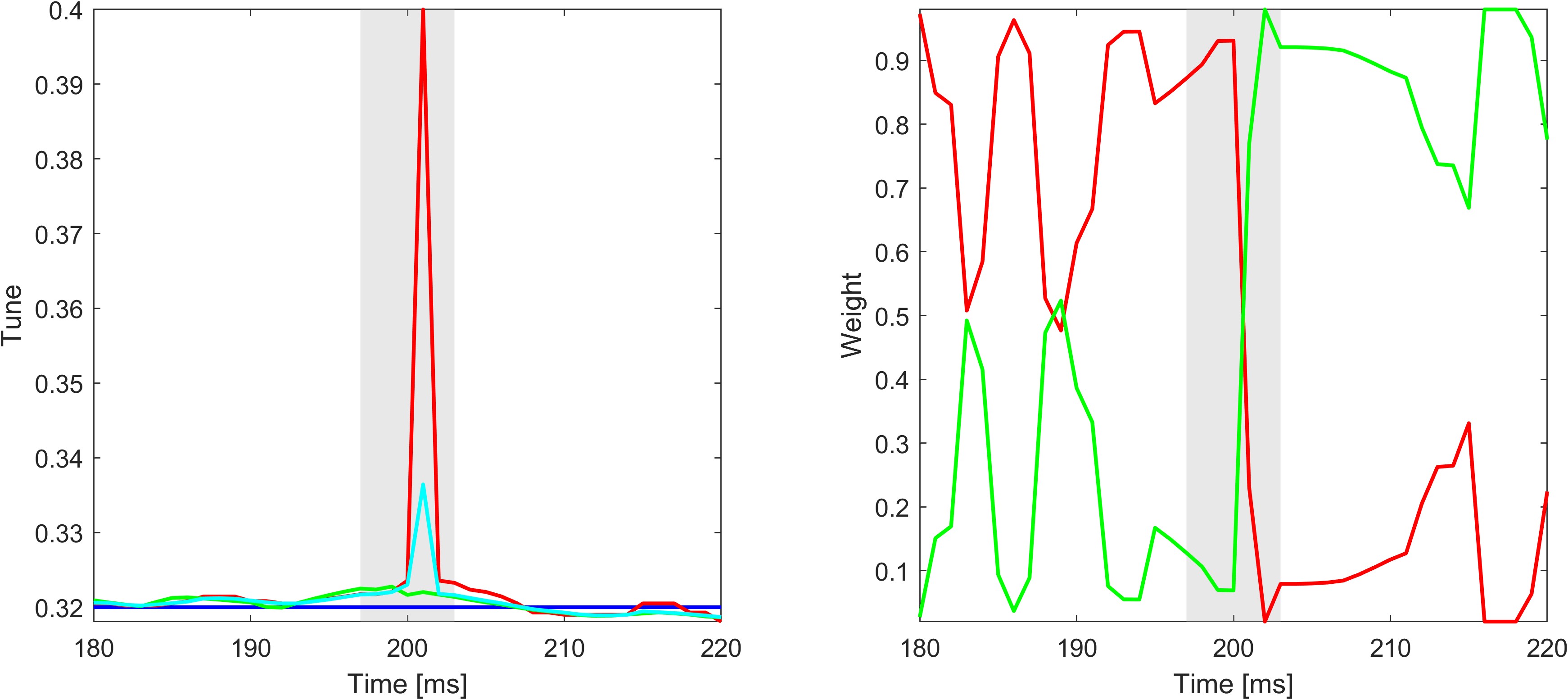}
    \caption{Fig.\ref{fig:amsf}(a) compares the filtered EMA tune (red line), filtered WLC tune (green line), actual tune (blue line), and predicted tune (cyan line) under an SNR of -20 dB with $q=0.68$. Fig.\ref{fig:amsf}(b) illustrates the weight variation of the EMA tune (red line) and WLC tune (green line). The convergence phase is not shown. In most cases, the Kalman filter tends to assign greater confidence to the EMA tune, as it is more stable and exhibits fewer fluctuations. However, when inevitable shot noise occurs, as indicated by the gray-shaded region in Fig.~\ref{fig:amsf}(a) and~\ref{fig:amsf}(b), the weight assigned to the EMA tune immediately decreases, prompting the filter to rely more on the WLC tune.}
    \label{fig:amsf}
\end{figure}

The adaptive multi-sensor fusion mechanism dynamically adjusts the weighting of the reference tune and measured tune, thereby further mitigating the impact of occasional shot noise. When one of the tune signals experiences fluctuations or enters an unstable state, its corresponding weight, \(w_i\), automatically decreases, indicating that the Kalman filter assigns greater confidence to the data from the other sensor. This dynamic adjustment ensures that the predicted results remain relatively stable, exhibiting lower bias (which enhances accuracy) and reduced standard deviation (which improves robustness), compared to only using online median filter solely for EMA tune or WLC tune. A comparison of the filtered EMA tune, filtered WLC tune, actual tune, and predicted tune using adaptive multi-sensor fusion is presented in Fig.~\ref{fig:amsf}.

\subsection{Results Post-Processing\label{subsec:postprocessing}}

After acquiring \( q_\text{pred} \), which is the final output of the tune measurement, a post-processing procedure is applied, particularly in cases where the BPM cannot measure the betatron tune across all energy regions or frequency ranges due to bandwidth limitations. Consequently, a validation step is necessary.

\begin{figure}
    \centering
    \includegraphics[width=0.5\hsize]{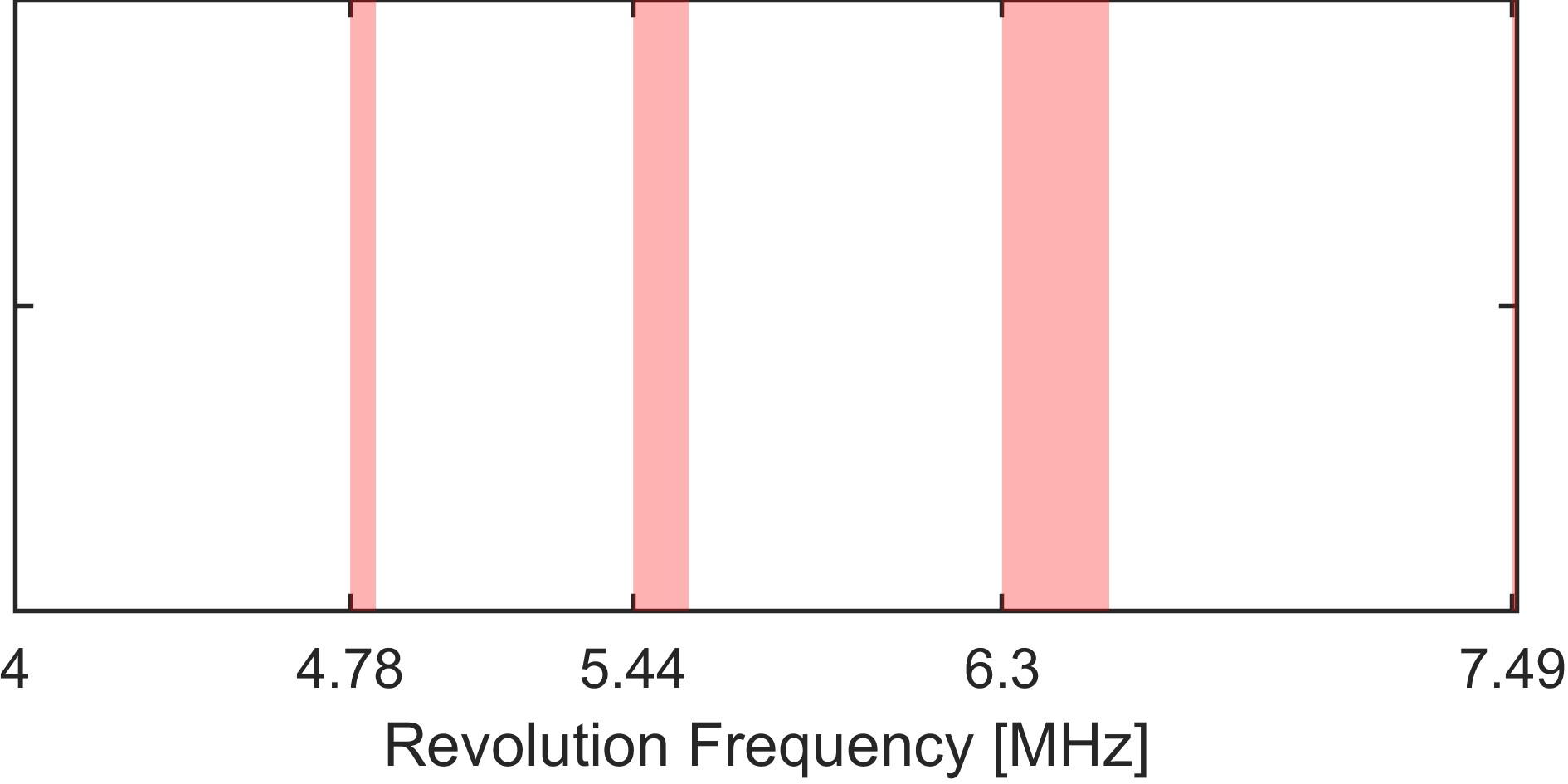}
    \caption{The red-shaded area represents the frequency region not covered by the BPM at \( q = 0.68 \). During operation, as the revolution frequency and tune vary dynamically, the uncovered region changes accordingly. This necessitates the post-processing procedure before outputting \( q_\text{pred} \) each time.}
    \label{fig:postprocessing}
\end{figure}

The bandwidth of the BPM is denoted as \( \text{BW}_{BPM} \), and its operating frequency is represented as \( f_c \). The nearest harmonic to \( f_c \) is then determined using:
\begin{align}
    n = \text{round} \left( \frac{f_c}{f_0} \right).
\end{align}
Subsequently, we evaluate whether the lower or upper transverse sideband of this harmonic falls within the BPM's bandwidth. If it does not, an \textbf{unreliable} flag is transmitted alongside the final output result to the upper-level control system, alerting the control room and other subsystems to potential inaccuracies. An example illustrating the covered and uncovered revolution frequency regions for \( q = 0.68 \) is shown in Fig.~\ref{fig:postprocessing}.

A latency compensation procedure is then applied by adjusting the time stamp of the results forward. This is necessary because the mapping, EMA and median filtering introduce system latency. This presents a trade-off between real-time performance and measurement accuracy. Prioritizing real-time performance entirely would make the system more susceptible to shot noise in low SNR environments, potentially leading to inaccurate tune measurements.

\subsection{System Implementation Architecture\label{subsec:workflow}}

The implementation architecture of the tune measurement system is outlined as follows:

\textbf{1. Data Preprocessing:}
\begin{enumerate}
    % \item[1.1] The FPGA generates an ADC-driven clock and a sampling clock with identical frequencies but distinct phases, calibrated according to the ADC properties.
    \item[1.1] The analog signal, obtained from a hybrid that outputs the differential of the input signals, is subsequently sampled and processed by a polyphase decimation filter that performs digital bandpass filtering to reduce the quantization noise power within the detector's bandwidth.
    \item[1.2] Two ping-pong FIFO buffers are implemented to manage the continuous incoming ADC data, ensuring no loss of critical data during FIFO read operations.
    \item[1.3] Based on the operating conditions of SAPT, an appropriate STFT window size is determined; these conditions can either be provided by the control room or determined by the FPGA itself based on recent frequency variations.
    \item[1.4] The unwanted coherent signal is removed using an RMS fitting method as described in Section~\ref{subsubsec:coherent_exclusion}.
\end{enumerate}

\textbf{2. Tune Measurement Procedure:}
\begin{enumerate}
    \item[2.1] Following preprocessing, the data are reorganized into multiple batches. The number of frequency bins per batch is computed based on the preconfigured minimal frequency resolution and remains constant throughout subsequent processing.
    \item[2.2] The PSD of each batch undergoes calculation, interpolating, smoothing and mapping operations, yielding $P_t$, which represents the PSD at time step $t$. Subsequently, a fixed frequency region encompassing the expected tune shift range is extracted for detailed analysis.
    \item[2.3] The EMA is updated using $P_t$; subsequently, the global maximum is identified and appended to the list $L_{\text{EMA}}$. The EMA tune, $q_{\text{EMA}}$, is computed as the median value of the last $N$ elements of $L_{\text{EMA}}$ (where $N$ denotes an optimally selected window size) to mitigate perturbations induced by shot noise.
    \item[2.4] The calculated reference tune $q_{\text{ref}}$ using Eq.~\ref{eqn:q_ref} serves as the ground truth, while the WLC tune, $q_{\text{WLC}}$, is determined by identifying the local maximum with highest confidence using the WLC algorithm. The resulting WLC tune is subsequently processed through an online median filter to enhance system robustness.
    \item[2.5] $q_{\text{EMA}}$ and $q_{\text{WLC}}$ are integrated via a Kalman filter algorithm to generate the final predicted tune, $q_{\text{pred}}$.
\end{enumerate}

\textbf{3. Post-Processing of Tune Measurement Results}
\begin{enumerate}
    \item[3.1] Given \(q_{\text{pred}}\) and the current revolution frequency \(f_0\), the system determines whether the combination of \(q_{\text{pred}}\) and \(f_0\) falls within the BPM’s bandwidth. If it does not, an \textbf{unreliable} flag is transmitted.
    \item[3.2] The time stamp of \(q_{\text{pred}}\) is adjusted forward to compensate for latency.
\end{enumerate}

Upon completion and after calibration and testing, the system will operate in conjunction with SAPT, delivering accurate real-time tune measurements across a wide range of energy levels, regardless of whether the beam is bunched or coasting. A comprehensive evaluation of the algorithm, based on data obtained from beam dynamics simulations using Xsuite, is presented in Section~\ref{sec:experiments}.

\section{\label{sec:experiments}Experiments}

The experiments were designed to evaluate and compare the performance of the proposed betatron tune measurement method with the conventional peak-detection algorithm~\cite{betz2017bunched} under SAPT conditions. To achieve this, SAPT design parameters were employed in a beam dynamics-based macro-particle simulation, where the revolution frequency was either linearly increased from 4 MHz to 7.5 MHz or held constant, with various types of tune variation introduced and an STFT time window applied. The simulated data were subsequently combined with real noise, acquired without beam, to emulate realistic measurement scenarios. In this section, multiple potential application scenarios are considered and analyzed. The analysis compares three key metrics for both the proposed and conventional algorithms: (1) The average absolute error of the measured tune relative to the nominal fractional tune, denoted as \( \mu \), which quantifies accuracy; (2) the standard deviation of the measured tune, denoted as \( \sigma \), which represents stability; and (3) the percentage of measured tunes falling within \( q \pm 0.01 \) and \( q \pm 0.001 \), denoted as \( P_{q\pm0.01} \) and \( P_{q\pm0.001} \), respectively, which evaluate compliance with design requirements. 
By systematically varying one parameter while keeping the others constant, this analysis provides a comprehensive evaluation of the proposed method’s performance relative to the conventional peak-detection algorithm, highlighting its advantages in accuracy and stability under challenging conditions.

All tune measurement results are obtained through online processing by inputting time-domain simulation data in 1 ms segments to emulate real-world application scenarios. The BPM operates at 36 MHz with an extended bandwidth of 8 MHz to ensure that at least one sideband is covered across all revolution frequencies in the ramping scenario, and at 39 MHz with a reduced bandwidth of 2 MHz to ensure that only a single sideband is captured when evaluating performance under tune shifts.

\subsection{Scenario 1: Performance under Ramping and Tune Shift\label{subsec:s1}}

Assuming the BPM has sufficient bandwidth to capture at least one sideband in each energy region, the revolution frequency and betatron tune of SAPT may not remain constant during operation. The worst-case scenario arises when both parameters vary rapidly, restricting the available sampling time. Given that ADC acquisition operates at a fixed constant sampling rate, extended sampling durations inherently encompass broader frequency ranges, potentially introducing spectral aliasing and leakage phenomena that significantly compromise tune measurement accuracy.

This section evaluates the performance of the proposed algorithm in comparison to the conventional peak-detection algorithm. First, either the revolution frequency or the betatron tune is varied while keeping the other parameter fixed under an SNR condition of \(-20\) dB. Subsequently, both the revolution frequency and the betatron tune are varied simultaneously to assess the algorithm's performance under more complex and challenging conditions.

\subsubsection{Ramping\label{s1.1}}

\begin{figure}
    \centering
    \includegraphics[width=0.5\hsize]{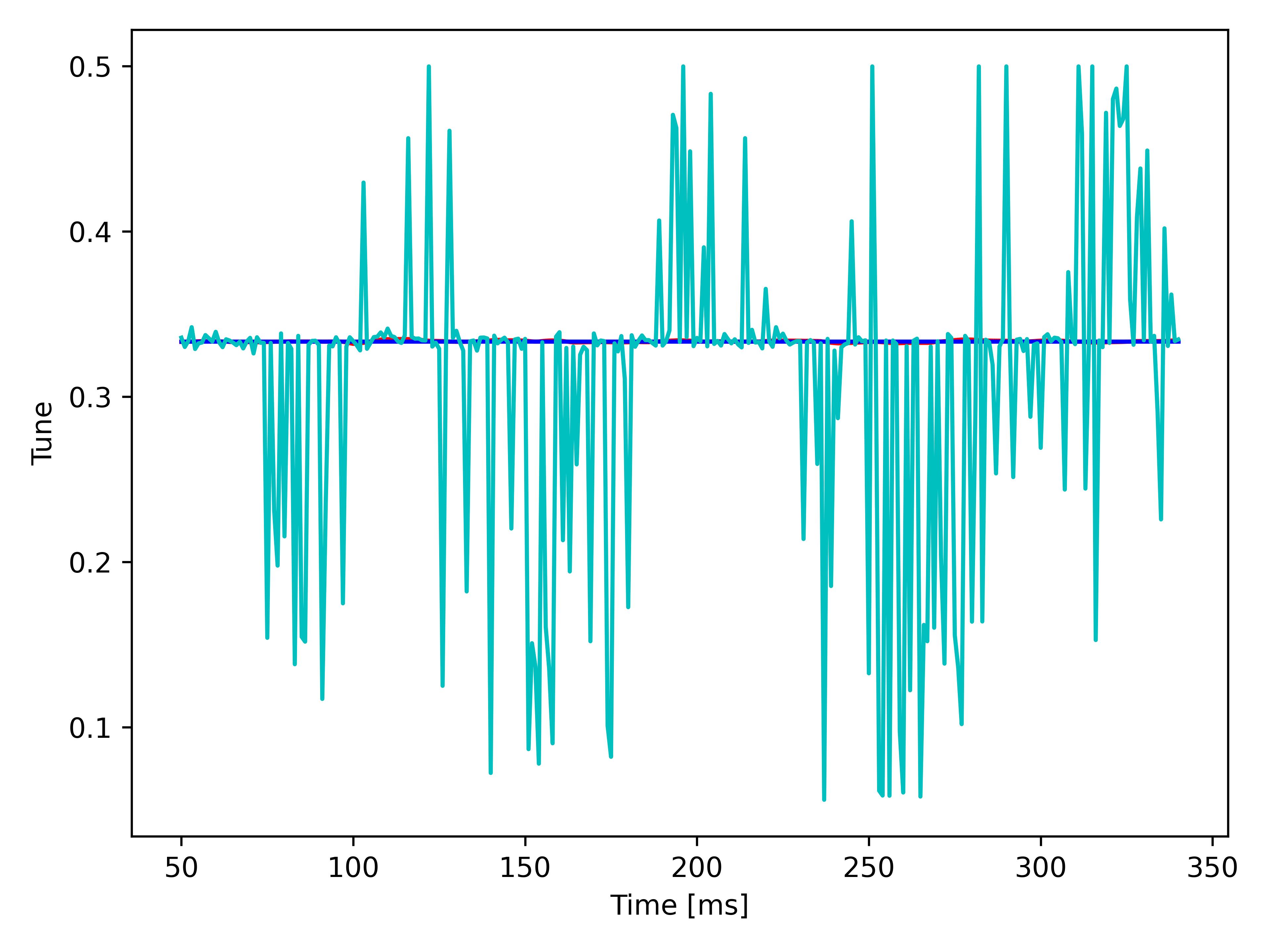}
    \caption{Comparison of tune measurement results between the proposed algorithm (red line) and the conventional peak-detection algorithm (cyan line), with the actual tune shown as the blue line. The figure demonstrates that the conventional peak-detection algorithm fails to accurately identify the betatron tune in low-SNR environments.}
    \label{fig:s1.1}
\end{figure}

The revolution frequency increases from 4 MHz to 7.5 MHz while ramping, as depicted in Fig.~\ref{fig:ramping}, while the betatron tune maintains a constant value of $q = 0.667$ under SNR conditions of $-20$ dB. The STFT time window is configured to 1 ms to accommodate the rapid increase in revolution frequency and to minimize spectral aliasing, leakage, and measurement bias. The comparative tune measurement results are presented in Fig.~\ref{fig:s1.1}, while the corresponding statistical parameters $\mu$, $\sigma$, $P_{q\pm0.01}$, and $P_{q\pm0.001}$ for both the proposed method and the conventional peak-detection algorithm are tabulated in Table~\ref{tab:s1.1}. The conventional peak-detection algorithm demonstrates significant limitations in accurately measuring the betatron tune under low SNR conditions. A comprehensive performance comparison between the proposed algorithm and the conventional peak-detection algorithm is presented in Section~\ref{subsec:s4}. Consequently, subsequent experiments focus exclusively on graphical representations of the proposed algorithm's performance, while numerical comparisons between both algorithms are presented in tabular format.

\begin{table}
\caption{\label{tab:s1.1}%
Performance metrics comparison between the proposed algorithm and conventional peak-detection algorithm under SNR conditions of -20 dB.
}
\centering
\begin{tabular}{lcccc}
\toprule
\textbf{Method} & \boldmath$\mu$ & \boldmath$\sigma$ & \boldmath$P_{q \pm 0.001}$ & \boldmath$P_{q \pm 0.01}$ \\
\midrule
Peak Detection
& 0.0229 & 0.0588 & 30.41\% & 84.79\%  \\
Proposed Algorithm
& 0.0007 & 0.0007 & 79.73\% & 100.00\%  \\
\bottomrule
\end{tabular}
\end{table}

\subsubsection{Tune Shift}

\begin{figure}
    \centering
    \includegraphics[width=\hsize]{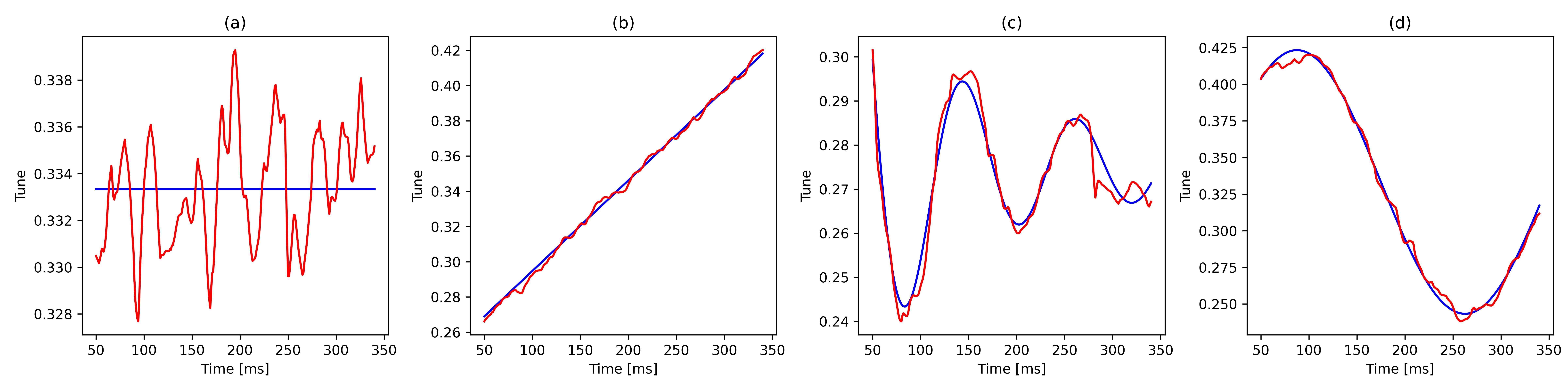}
    \caption{Comparison of tune measurement results between the proposed algorithm (red line) and the nominal tune (blue line) for four different types of tune shifts.}
    \label{fig:s1.2}
\end{figure}

The revolution frequency will be maintained at a constant value of 7.5 MHz, while the tune is varied in different ways. The performance of the proposed algorithm under an SNR of -20 dB is illustrated in Fig.~\ref{fig:s1.2}, while the numerical comparison between the proposed algorithm and the conventional peak-detection algorithm is presented in Table.~\ref{tab:s1.2}.

\begin{table}
\caption{\label{tab:s1.2}%
Performance metrics comparison between the proposed algorithm and conventional peak-detection algorithm under SNR conditions of -20 dB.
}
\centering
\begin{tabular}{lcccc}
\toprule
\textbf{Method} & \boldmath$\mu$ & \boldmath$\sigma$ & \boldmath$P_{q_\pm0.001}$ & \boldmath$P_{q\pm0.01}$ \\
\midrule
Peak Detection
& 0.0775 & 0.0929 & 5.58\% & 37.71\%  \\
Proposed Algorithm
& 0.0023 & 0.0019 & 29.21\% & 99.83\%  \\
\bottomrule
\end{tabular}
\end{table}

\subsubsection{Ramping and Tune Shift}

\begin{figure}
    \centering
    \includegraphics[width=\hsize]{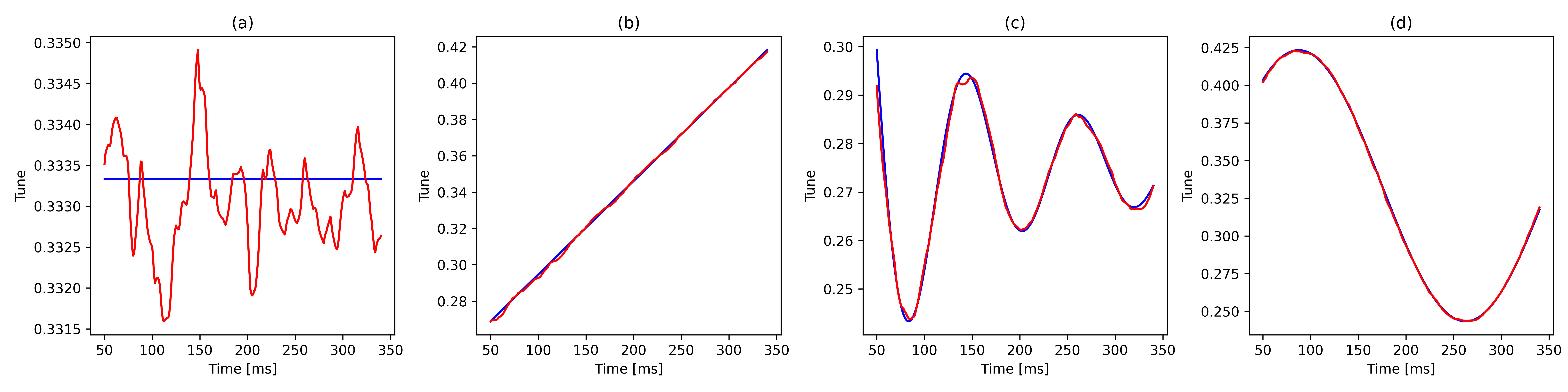}
    \caption{Comparison of tune measurement results between the proposed algorithm (red line), with the nominal tune shown as the blue line.}
    \label{fig:s1.3}
\end{figure}

During ramping, the simultaneous increase in revolution frequency and tune shift limits the available sampling time. The STFT time window size is set to 1 ms, as described in Section~\ref{s1.1}. The performance and numerical comparison are presented in Fig.~\ref{fig:s1.3} and Table~\ref{tab:s1.3}.

\begin{table}
\caption{\label{tab:s1.3}%
Performance metrics comparison between the proposed algorithm and conventional peak-detection algorithm under SNR conditions of -20 dB.}
\centering
\begin{tabular}{lcccc}
\toprule
\textbf{Method} & \boldmath$\mu$ & \boldmath$\sigma$ & \boldmath$P_{q\pm0.001}$ & \boldmath$P_{q\pm0.01}$ \\
\midrule
Peak Detection
& 0.0229 & 0.0588 & 30.41\% & 84.79\%  \\
Proposed Algorithm
& 0.0007 & 0.0007 & 79.73\% & 100.00\%  \\
\bottomrule
\end{tabular}
\end{table}

\subsection{Scenario 2: Performance under Data Contamination and Signal Loss}

\begin{figure}
    \centering
    \includegraphics[width=\hsize]{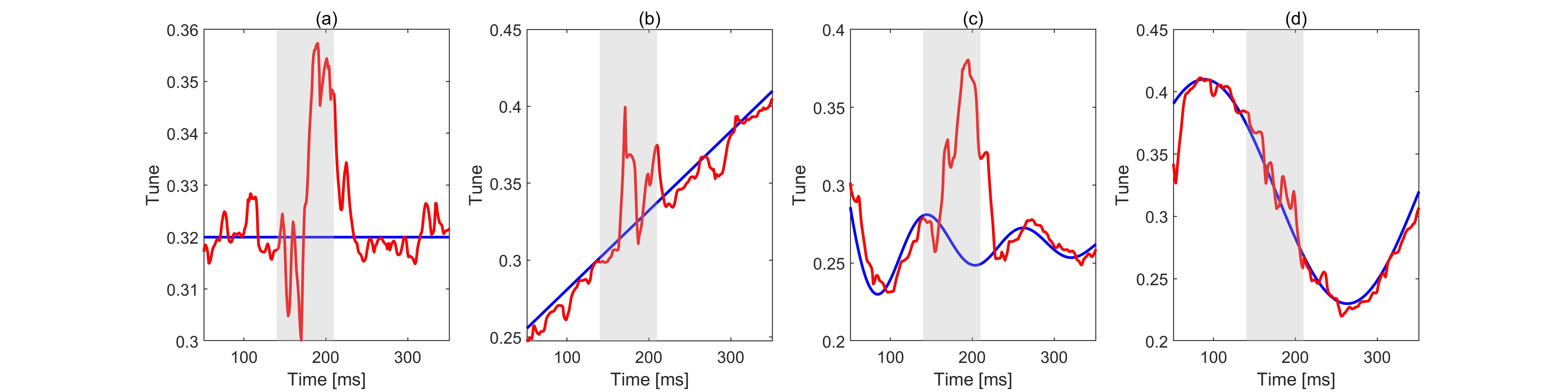}
    \caption{Comparison of performance between the proposed algorithm (red line) and the nominal tune (blue line) under data contamination and signal loss. The gray-shaded area represents the region where no transverse sideband is present, and only noise exists. After exiting the contamination zone, the algorithm requires approximately 20 ms to converge to the actual tune.}
    \label{fig:s2}
\end{figure}

During daily operation, unexpected disturbances or hardware limitations may introduce strong noise into the ADC-acquired data or result in the complete absence of transverse signal content. The latter scenario occurs in SAPT because the bandwidth of the developing detector is insufficient to cover all energy regions during ramping or extraction. In this case, the transverse sideband may not appear in the spectra, leaving only noise. Consequently, it is crucial to assess the algorithm's ability to converge to the actual tune value after being affected by disturbances. The performance under data contamination or signal loss, with a revolution frequency of 7.5 MHz, an SNR of \(-20\) dB, and a 1 ms STFT time window, is illustrated in Fig.~\ref{fig:s2}. Given a 1 ms STFT time window, the measured results converge to the actual value in less than 50 ms. 

% \subsection{Scenario 3: Performance under Relatively Stable Revolution Frequency\label{subsec:s2}}

% If the revolution frequency remains stable, it is practical to use a longer sampling time, such as 5 ms, to achieve more precise tune measurements. This approach also mitigates the influence of shot noise in low-SNR environments. The performance of the proposed algorithm, along with numerical metrics, is compared with the conventional peak-detection algorithm in Fig.~\ref{fig:s3} and Table~\ref{tab:s3}.

% \begin{figure}
%     \centering
%     \includegraphics[width=\hsize]{s3_comparison.jpg}
%     \caption{Comparison of the performance between the proposed algorithm (red line) and the nominal tune (blue line) under a relatively stable revolution frequency and a longer sampling time. Occasional shot noise is no longer present due to the summation of more batches, which increases the SNR.}
%     \label{fig:s3}
% \end{figure}

% \begin{table}
% \caption{\label{tab:s3}%
% Performance metrics comparison between the proposed algorithm and conventional peak-detection algorithm under SNR conditions of -20 dB.}
% \centering
% \begin{tabular}{lcccc}
% \toprule
% \textbf{Method} & \boldmath$\mu$ & \boldmath$\sigma$ & \boldmath$P_{q \pm 0.001}$ & \boldmath$P_{q \pm 0.01}$\\
% \midrule
% Peak Detection
% & 0.0065 & 0.0211 & 20.53\% & 92.17\%  \\
% Proposed Algorithm
% & 0.0010 & 0.0008 & 62.23\% & 98.37\% \\
% \bottomrule
% \end{tabular}
% \end{table}

\subsection{Scenario 3: Performance under the Absence of Coherent Tune\label{subsec:s4}}

\begin{figure}
    \centering
    \includegraphics[width=\hsize]{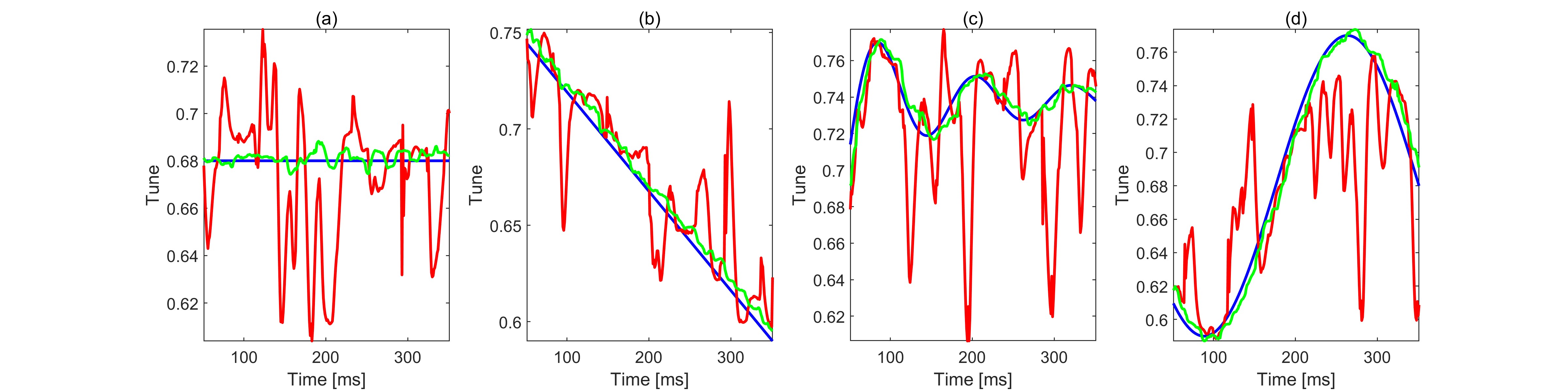}
    \caption{Comparison of the performance of the proposed algorithm under -20 dB SNR (red line) and -15 dB SNR (green line), with the actual tune shown as the blue line. The absence of a coherent signal reduces the power of the transverse sideband, thereby increasing the minimum SNR required for accurate tune measurement.}
    \label{fig:s4}
\end{figure}

During SAPT operation, obtaining a coherent tune signal from BPM data is challenging. Therefore, it is essential to evaluate the algorithm's performance in the absence of a coherent tune signal. The noise amplitude is calculated based on the signal containing a coherent component, after which the coherent tune signal content is removed using the method described in Section~\ref{subsubsec:coherent_exclusion}. The remaining signal is then combined with noise. Experimental results indicate that the absence of a coherent tune signal slightly increases the minimum SNR required for accurate tune measurement. However, at an SNR of -15 dB, the algorithm still demonstrates good accuracy with relatively low error, as shown in Fig.~\ref{fig:s4} and Table~\ref{tab:s4}.

\begin{table}
\caption{\label{tab:s4}%
Performance metrics comparison between the proposed algorithm and conventional peak-detection algorithm under SNR conditions of -15 dB.}
\centering
\begin{tabular}{lcccc}
\toprule
\textbf{Method} & \boldmath$\mu$ & \boldmath$\sigma$ & \boldmath$P_{q \pm 0.001}$ & \boldmath$P_{q \pm 0.01}$\\
\midrule
Peak Detection
& 0.0374 & 0.0436 & 3.98\% & 36.85\%  \\
Proposed Algorithm
& 0.0057 & 0.0044 & 11.45\% & 85.76\%  \\
Proposed Algorithm (latency compensated)
& 0.0042 & 0.0034 & 15.94\% & 94.72\%  \\
\bottomrule
\end{tabular}
\end{table}

\subsection{Scenario 4: Performance under Different SNR\label{subsec:s5}}

\begin{figure}
    \centering
    \includegraphics[width=\hsize]{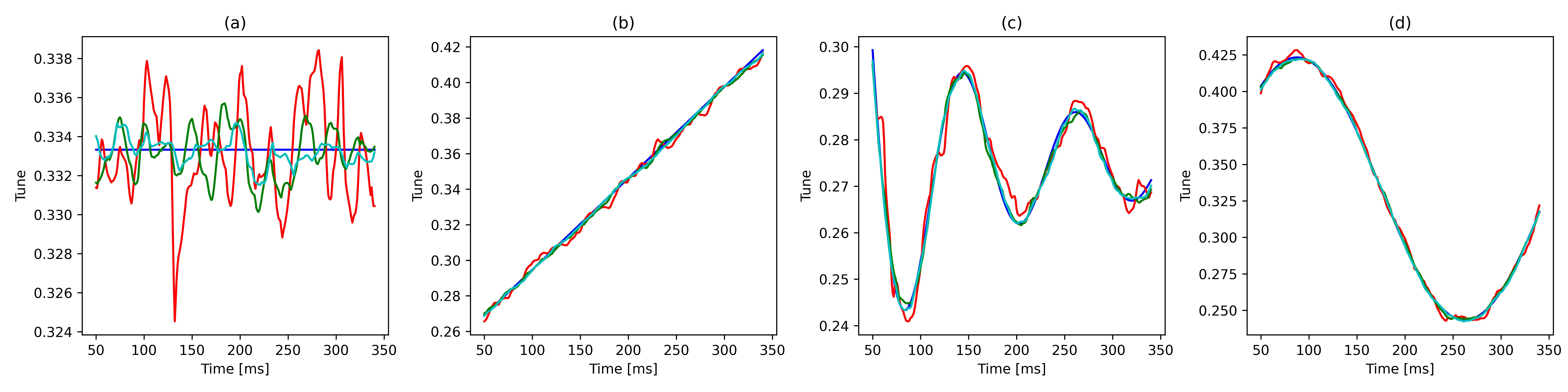}
    \caption{Comparison of the performance of the proposed algorithm under -20 dB SNR (red line), -15 dB SNR (green line) and -10 dB SNR (cyan line), with the nominal tune shown as the blue line.}
    \label{fig:s5}
\end{figure}

The SNR of the acquired signal is influenced by multiple factors. To ensure the generalization and accuracy of the proposed method, it is essential to evaluate its performance across different SNR values. The comparison between the proposed algorithm and the conventional peak-detection algorithm is illustrated in Fig.~\ref{fig:s5}, while the numerical results are presented in Table~\ref{tab:s5}.

\begin{table}
\caption{\label{tab:s5}%
Performance metrics comparison between the proposed algorithm and conventional peak-detection algorithm under SNR conditions of -20 dB, -15 dB and -10 dB.}
\centering
\begin{tabular}{llcccc}
\toprule
\textbf{SNR} & \textbf{Method} & \boldmath$\mu$ & \boldmath$\sigma$ & \boldmath$P_{q \pm 0.001}$ & \boldmath$P_{q \pm 0.01}$\\
\midrule
\multirow{2}{*}{-20 dB} 
& Peak Detection
& 0.0764 & 0.0900 & 4.73\% & 37.29\%  \\
& Proposed Algorithm
& 0.0022 & 0.0017 & 27.41\% & 99.66\%  \\
\midrule
\multirow{2}{*}{-15 dB} 
& Peak Detection
& 0.0087 & 0.0302 & 21.22\% & 91.75\%  \\
& Proposed Algorithm
& 0.0010 & 0.0007 & 54.47\% & 100.00\%  \\
\midrule
\multirow{2}{*}{-10 dB} 
& Peak Detection
& 0.0022 & 0.0018 & 30.76\% & 99.91\%  \\
& Proposed Algorithm
& 0.0007 & 0.0006 & 74.91\% & 100.00\%  \\
\bottomrule
\end{tabular}
\end{table}

\subsection{Scenario 5: Performance under Tune Jump\label{subsec:s6}}

During operation, tune jumps may occur. These abrupt changes in betatron tune require the tune measurement system to detect such events, rather than classify them as outliers, and to converge to the actual value as quickly as possible while maintaining accuracy and precision. 

To evaluate the proposed algorithm’s ability to handle this scenario, experiments were conducted under an SNR of -20 dB with an STFT time window of 1 ms to assess the algorithm’s performance under the worst conditions in which it can operate. 

\begin{figure}
    \centering
    \includegraphics[width=0.5\hsize]{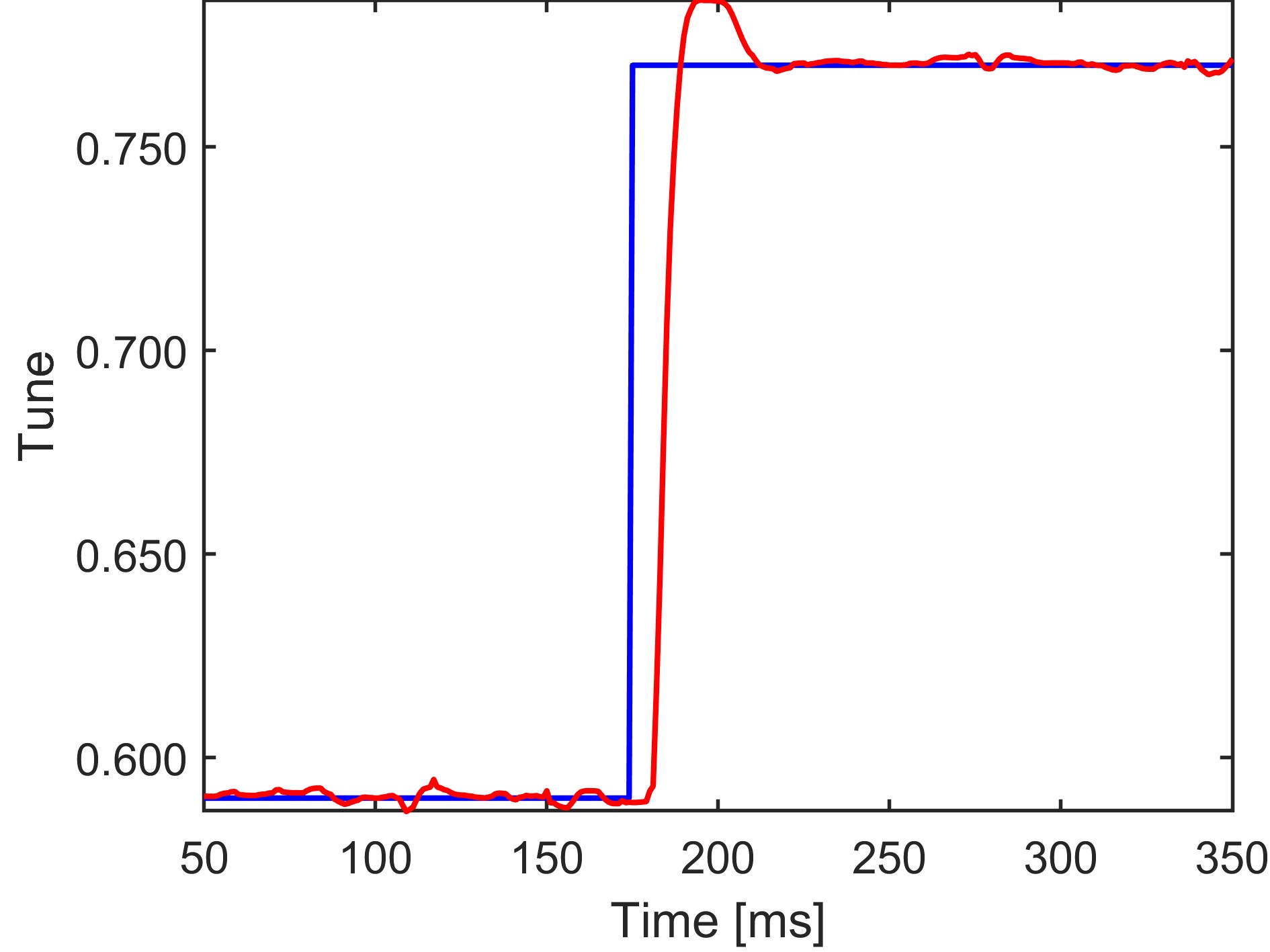}
    \caption{Comparison of the proposed algorithm's performance under a tune jump scenario. The measured results are represented by the red line, while the nominal tune is depicted by the blue line. Following the tune jump, after a brief latency, the measured results rapidly converge to the actual tune value and remain stable. The entire process takes less than 50 ms.}
    \label{fig:s6}
\end{figure}

The performance results are shown in Fig.~\ref{fig:s6}.

\subsection{Scenario 6: Performance under Drifting\label{subsec:s2}}

\begin{figure}
    \centering
    \includegraphics[width=\hsize]{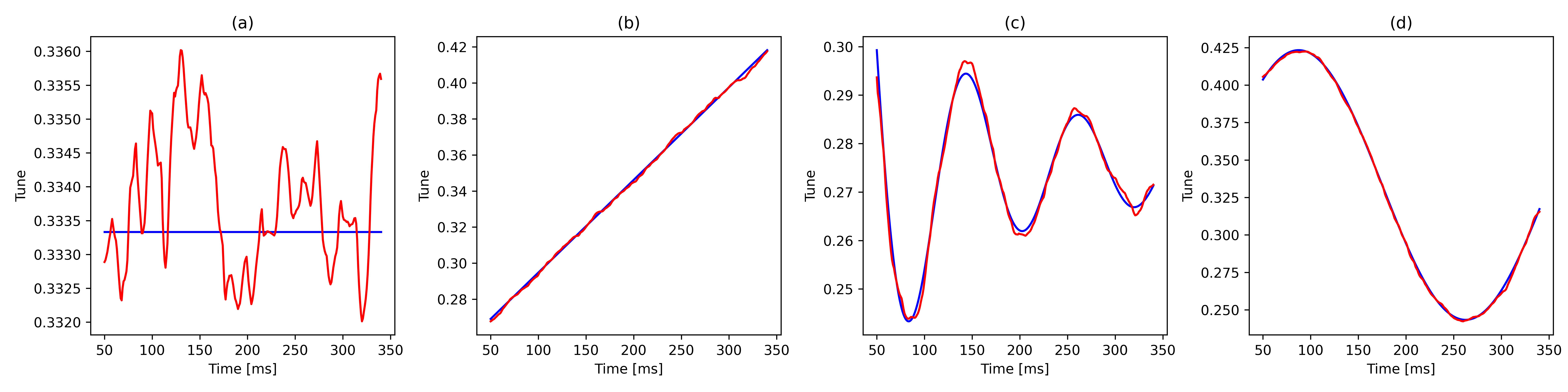}
    \caption{Comparison of the performance of the proposed algorithm (red line) and the nominal tune (blue line) under drifting beam conditions.}
    \label{fig:s7}
\end{figure}

Tune measurements must be performed during either bunching or drifting. Therefore, it is essential to evaluate the performance of the proposed algorithm under drifting conditions. The performance of the proposed algorithm, along with numerical metrics, is compared against that of the conventional peak-detection algorithm, as illustrated in Fig.\ref{fig:s7} and Table\ref{tab:s7}.

\begin{table}
\caption{\label{tab:s7}%
Performance metrics comparison between the proposed algorithm and conventional peak-detection algorithm under SNR conditions of -20 dB.}
\centering
\begin{tabular}{lcccc}
\toprule
\textbf{Method} & \boldmath$\mu$ & \boldmath$\sigma$ & \boldmath$P_{q \pm 0.001}$ & \boldmath$P_{q \pm 0.01}$\\
\midrule
Peak Detection
& 0.0547 & 0.0868 & 15.21\% & 64.09\%  \\
Proposed Algorithm
& 0.0009 & 0.0007 & 63.32\% & 100.00\% \\
\bottomrule
\end{tabular}
\end{table}

\subsection{Results Discussion}

The experimental results demonstrate the superior performance of the proposed betatron tune measurement algorithm compared to the conventional peak-detection method under various challenging conditions. The findings are summarized as follows:

\subsubsection{Performance under Ramping and Tune Shift}
\begin{itemize}
    \item The proposed algorithm effectively tracks betatron tune variations even when the revolution frequency changes at a rate of approximately 10 MHz/s while ramping.
    \item Under an SNR of -20 dB, the conventional peak-detection algorithm exhibits reduced accuracy due to strong noise spectrum interference, whereas the proposed algorithm maintains significantly lower error and higher stability.
    \item When both the revolution frequency and tune shift occur simultaneously, the proposed method continues to provide reliable measurements, demonstrating robustness in complex scenarios.
\end{itemize}

\subsubsection{Performance under Data Contamination and Signal Loss}
\begin{itemize}
    \item The proposed algorithm successfully converges to the actual tune value after disturbances, with convergence occurring within 50 ms.
\end{itemize}

% \subsubsection{Performance under Relatively Stable Revolution Frequency}
% \begin{itemize}
%     \item With a longer sampling duration, the proposed algorithm achieves enhanced precision, benefiting from improved frequency resolution and reduced noise influence.
%     \item The statistical results indicate that increasing the sampling duration significantly improves measurement stability while maintaining responsiveness to tune variations.
% \end{itemize}

\subsubsection{Performance under the Absence of Coherent Tune}
\begin{itemize}
    \item The removal of the coherent signal slightly increases the minimum required SNR for accurate tune measurements.
    \item Nevertheless, at an SNR of -15 dB, the proposed algorithm maintains relatively low measurement error, confirming its applicability even in the absence of a strong coherent tune signal.
\end{itemize}

\subsubsection{Performance under Different SNR Conditions}
\begin{itemize}
    \item As the SNR improves from -20 dB to -10 dB, the accuracy and precision of the proposed method steadily increase.
    \item The algorithm consistently outperforms the conventional peak-detection method, particularly in low-SNR environments, where traditional methods exhibit severe degradation.
\end{itemize}

\subsubsection{Performance under Tune Jumps}
\begin{itemize}
    \item The algorithm successfully detects abrupt tune jumps and rapidly converges to the actual tune value within 50 ms.
    \item The proposed method correctly identifies these jumps as genuine changes and maintains accuracy and stability after convergence.
\end{itemize}

\subsubsection{Performance under Drifting}
\begin{itemize}
    \item The proposed algorithm effectively tracks betatron tune variations under drifting beam conditions at an SNR of -20 dB.
\end{itemize}

\subsection{Summary of Experimental Findings}

The experimental results confirm that the proposed betatron tune measurement algorithm offers superior accuracy, stability, and robustness across diverse operational conditions. Its ability to handle rapid frequency variations, noise contamination, signal loss, and abrupt tune jumps makes it a reliable alternative to conventional methods, particularly in low-SNR environments.

\section{Limitations and Future Work\label{sec:limitations}}

The resonant stripline BPM for betatron tune measurement is still under development, making it currently infeasible to detect the Schottky signal and validate the proposed algorithm on the existing SAPT facility. Additionally, the BPM’s performance remains uncertain. To address this limitation, the proposed algorithm was evaluated through macro-particle simulations incorporating realistic beam dynamics models based on SAPT design parameters and actual noise, ensuring that the simulated conditions closely resemble real experimental environments. The validation under an SNR as low as -20 dB, as discussed in Section~\ref{sec:experiments}, was conducted to account for potential BPM performance constraints in practical applications. 

Upon completion of BPM development, manufacturing, and installation at SAPT or an alternative proton therapy synchrotron facility, an FPGA implementation of the proposed algorithm will undergo comprehensive validation using authentic experimental data. The evaluation will include direct comparisons with conventional tune measurement methods to assess accuracy and robustness in an operational setting. If BPM development is delayed, alternative validation strategies, such as testing on existing BPM systems with partial implementation, will be explored to provide additional experimental support.

\section{\label{sec:conclusion}Conclusion}

In this paper, we proposed a novel Schottky diagnostics-based method for real-time betatron tune measurement in SAPT. The method addresses critical challenges such as low SNR, varying revolution frequency, and fluctuating betatron tune. By leveraging macro-particle beam-dynamics simulations and incorporating real-world noise, we demonstrated the method's capability to extract transverse Schottky signals from noisy environments. The proposed approach utilizes STFT combined with advanced smoothing and signal processing techniques to achieve accurate betatron tune measurements under a wide range of experimental conditions.

Experimental results demonstrate that the proposed method significantly outperforms the conventional peak-detection algorithm in terms of precision, accuracy, and robustness. The method achieves an average absolute error (\( \mu \)) relative to the nominal fractional tune of less than \( 0.01 \), with a low standard deviation (\( \sigma \)), thereby meeting the stringent design requirements for high-accuracy tune diagnostics. These results highlight the method's general applicability and stability, even under challenging operational scenarios such as rapid frequency ramping, tune shifts, and low SNR conditions.

This study lays the groundwork for optimizing betatron tune diagnostics in ramping synchrotrons and advancing diagnostic techniques for applications such as proton therapy. Future work will focus on expanding the operational range of the proposed method and validating its applicability in other synchrotron facilities. Additionally, efforts will be directed toward further optimizing computational efficiency to enhance real-time performance in demanding environments.

To facilitate practical implementation, an FPGA-based online system for the proposed method is currently under development. This system aims to provide high-accuracy, real-time betatron tune measurements for SAPT, further enhancing the diagnostic capabilities of modern synchrotron facilities.

\bibliographystyle{JHEP}
\bibliography{main}

\end{document}